\documentclass[preprint]{aastex}

\begin{document} 

\title{The shock reprocessing model of electron acceleration in impulsive solar flares}
\author {Robert Selkowitz and Eric G. Blackman}
\affil{Dept. of Physics and Astronomy, and Laboratory for Laser Energetics, 
University of Rochester, Rochester NY, 14627} 

\begin{abstract}

We propose a new two-stage model for acceleration of electrons
in solar flares.  In the first stage, electrons are accelerated
stochastically in a post-reconnection turbulent downflow.
The second stage is the reprocessing of a subset of these electrons
as they pass through a  weakly compressive fast shock above
the apex of the closed flare loop on their way to the chromosphere.
We call this the "shock reprocessing" model. The model reproduces
the energy dependent arrival time delays observed
for both the pulsed and smooth components of impulsive solar
flare x-rays with physically reasonable parameters for the downflow
region. The model also predicts an emission site above the loop-top, as 
seen in the Masuda flare. The loop-top source distinguishes the shock
reprocessing model from previous models.
The model makes testable predictions for the energy
dependence of footpoint pulse strengths and the location and spectrum
of the loop-top emission, and can account for the observed soft-hard-soft
trend in the spectral evolution of footpoint emission.
Our model highlights the concept that reconnection is  
an acceleration environment rather than a single process.
Which combination of processes operate may depend on the initial conditions
that determine, for example, whether the reconnection downflow
is turbulent.  The shock reprocessing model comprises one such combination.

\bigskip
\end{abstract}

\section{Introduction}

An important observational constraint on the acceleration and transport processes in solar flares is imposed by X-ray arrival time 
delay measurements \citep{Aschwanden, A96a, A96b, A97, A98a, A98b, A99}.  It is observed that the 
non-thermal hard X-ray emission consists of two separable components, a smooth,
slowly varying background, and a pulsed, rapidly varying modulation.  Detailed studies of the temporal structure of the two 
components show that they possess several opposite characteristics:  Typically, the variations in the smooth component are 
first observed at 
lower energies, then at increasingly higher energy.  The arrival time delay between $25$keV and $150$ keV electrons is of order $5$s.  
Conversely, the pulses are observed first in the highest energies, with lower energy photons appearing later; the lag time for $25$keV
photons relative to $150$keV photons is of order $50$ms.  The trap-precipitation 
model of \citet{BrownMelrose} offers one possible explanation for the time delay measurements.  In this model, electrons are 
accelerated to high energy above the loop in the turbulent downflow and are then injected into the closed flare loop on
the pulse time scale.  
Some electrons are trapped between magnetic compressions in the loop before precipitating onto the chromosphere, 
while the remainder precipitate directly.  The two populations produce the smooth and pulsed components respectively.  

Although the trap precipitation model may plausibly account for the time delay observations in some flares, 
some significant uncertainties remain.  
While the needed injection may arise naturally as a feature of many acceleration models which are situated in the 
reconnection current sheet (i.e. \citet{Litvinenko}), it is not clear how injection would take place in models which
employ acceleration in the turbulent downflow region which may form between the reconnection sheet and the closed soft x-ray loops. In addition, 
  \citet{LarosaShore} argue that the pulsed structure can be produced by fluctuations in spectral hardness instead of a pulsed injection
of non-thermal electrons.  Furthermore, the model predicts that coronal X-ray 
emission emanates from
within the trap region.  The trap precipitation model places the trap, and thus this emission site, within the closed flare 
loop itself. This is seemingly at odds with at least some observations,
such as those of of \citet{Sui, Masuda, alexander} which 
reveal a compact non-thermal
source above the loop-top during the main phase of some flares. 
Another disadvantage of the trap precipitation
model is that it does not retain  temporal information of 
the acceleration process; such information is lost 
at the injection point. A different scenario in which the 
properties of the acceleration mechanism are more closely linked to the emission may provide more opportunity to test both the time 
delay mechanism and the acceleration mechanism observationally.  
We propose a new model called the ``shock reprocessing model'' as 
just such an alternative; it circumvents the loop injection problem, places the coronal emission above the loop-top, and retains some 
details of the acceleration mechanism in the footpoint emission.

The shock reprocessing model employs both second order stochastic Fermi acceleration
(STFA) and first order acceleration at fast shocks.  Electrons are accelerated via STFA in the turbulent downflow region located
below the reconnection point.  
Due to the fairly long acceleration time, typically a few seconds, variations on the eddy time, $\backsim 1$s, 
are smoothed out in the temporal spectrum of the STFA accelerated population.  STFA, operating on a time scale of a few seconds,  
produces the smooth component of the X-ray emission described in \citet{Aschwanden}.  As the flow impinges upon the flare loop,  
typical flow speeds are marginally super-magnetosonic (e.g. Blackman \& Field 1994).  
A weakly compressive fast shock forms.  
Because the flow is only marginally super-magnetosonic,
the shock can disappear and reform as the flow speed fluctuates on the eddy time.  This is a key feature of the shock reprocessing 
model. First order Fermi acceleration occurs at the fast shock and the transient nature of the shock can be responsible for the pulsed emission 
structure observed in impulsive solar flares.  
Furthermore, the shock is strongest above the apex of the loop, and diminishes toward
the wings.  The shock formation condition is typically met over a portion of the loop; as a result, only a fraction of the 
STFA accelerated electrons undergo the second phase of acceleration..

We discuss the details of the acceleration mechanisms 
in section 2.  In section 3 we determine the range of shock parameters 
required by the shock reprocessing model, as well as a simple model of shock formation.  
In particular, we calculate the compression ratio and spatial filling fraction of the shocks. In section 4 we discuss the role of 
cooling in the STFA region in matching the observed smooth component time delays and the 
presence of loop-top emission sites in a small fraction of Yohkoh flares, such as the Masuda flare \citep{Nature}.  
The distinct presence of these X-ray sources just above and exterior to the loop-top is a distinguishable feature of the shock 
reprocessing model when compared to the trap precipitation model.  We conclude in section 5.

\section{Flare properties and acceleration mechanisms}

Before examining the shock reprocessing model in depth, we briefly review models of flare morphology and 
some properties of STFA and shock Fermi acceleration which will later be employed.

\subsection{Flare properties}

It is well understood that solar flares are driven by magnetic reconnection in the solar corona (i.e., \citet{Priest}).  
The reconnected field lines relax into a closed loop structure evidenced by the soft x-ray loops which are filled with hot
plasma.  Hard x-rays are observed at the chromospheric footpoints of the closed loops as well coronal hard x-ray sources in some 
flare.  The acceleration mechanisms responsible for the hard-xray producing electrons are not yet fully known; candidates include
DC field acceleration \citep{Litvinenko}, stochastic acceleration \citep{Miller, chandran, selkblack}, and shock acceleration
\citep{Tsuneta2}, or combinations of these processes \citep{Kosugi}.  Further complicating the issue is 
that the downflow plasma between the reconnection point and the relaxed loops may either be laminar or turbulent.  Laminar
flow would favor stationary \citep{BrownMelrose}and collapsing \citet{karlicky} 
trap precipitation models of electron acceleration, which require an ordered field structure.  
Turbulent downflow would favor stochastic acceleration models, which require turbulent field structure.  

\begin{figure}
\epsscale{1.0}
\plotone{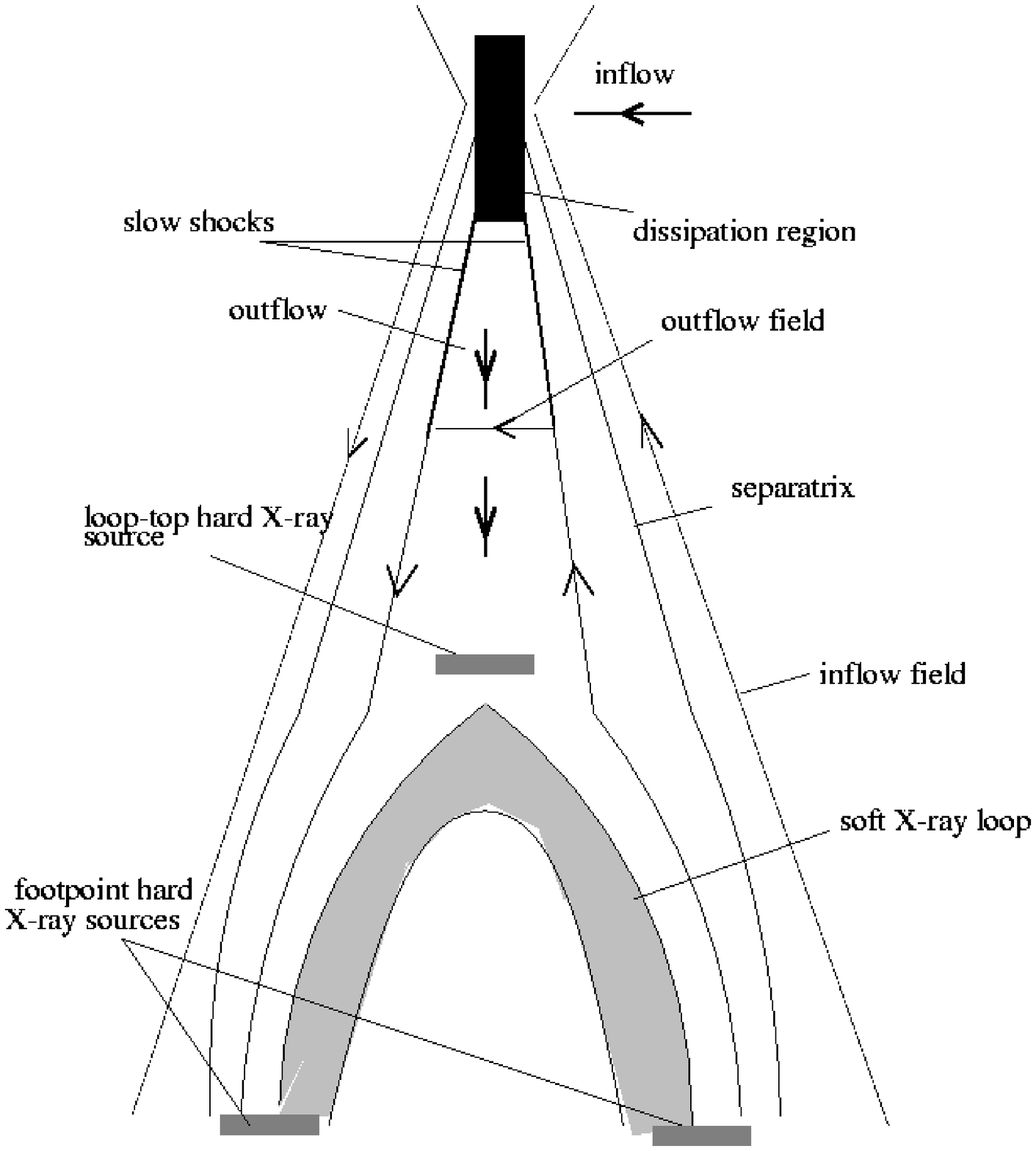}
\caption{The basic model of solar flare structure \citet{Blackman97}.  Reproduced by permission of the AAS.}
\label{cartoon}
\end{figure}

In addition, there are two possible directions of motion to consider 
for the evolution of the reconnection point.  The 
standard model (i.e. \citet{Priest}) focuses on the vertical direction of motion
in which the reconnection point moves upward through the corona over time.  As a result, 
the footpoint separation and height of the soft x-ray loop increases with time.  
However, recent observations with 
RHESSI indicate that at some flares have a ``zipper'' morphology; the reconnection evolves
laterally along an arcade structure at near constant height, 
with the soft x-ray loop following the reconnection point along the arcade at near constant 
footpoint separation and loop height \citep{NewBenz}.  
Regardless of the direction of motion of the reconnection point, a 2-D cross section of 
flare morphology is shown in figure \ref{cartoon}.  The region labelled outflow may be either turbulent or laminar, depending on 
the local Alfv\'en speed and the initial gradient in the downflow
across the downflow cross section (Chieuh \& Zweibel 1987).  
 The shock reprocessing model focuses primarily on the vertically evolution with 
turbulent downflows.

It is important to note that there is also 
a variety of coronal hard x-ray emission characteristics observed.  
The discovery event, dubbed the Masuda flare \citep{Nature},shows hard x-rays above the closed loop with 
a non-thermal kernel \citep{alexander}
as well as a very hot thermal region.  On the other hand, 
many coronal sources appear to be wholly thermal \citep{emslie}. 
Some are located inside the soft x-ray loops, not above \citep{veronig}.  
The variety of x-ray emission properties and reconnection morphologies implies that even
if all solar flares result from a basic reconnection environment and a soft x-ray loop, 
they are not explained within a single scenario of specific processes 
such as trap precipitation or shock reprocessing.
Key elements, such as laminar vs. turbulent downflow may determine the differences
between some classes of flares.
A necessary condition for turbulence to form in the 
reconnection outflow, super-Alfv\'enic downflow speeds, is also an approximate result of 
reconnection outflows \citep{Blackman94,karlicky}.  
Ultimately, models which consider flares with and without turbulent downflows 
need to be comparatively explored.

\subsection{Power law acceleration solely by STFA}

STFA is capable of accelerating solar flare electrons out of a thermal population \citep{Miller, LaRosa, chandran, selkblack} 
and up to high energies with a spectral shape that depends mainly on the mechanism which traps
electrons in the acceleration region.  In the reconnection outflow, trapping is provided by wave particle interactions which also 
serve as the pitch angle scattering necessary to sustain acceleration \citep{achterberg}.  Below a transition energy 
$E_t \backsim 10$keV, pitch angle scattering in solar flares occurs through interactions with whistler waves.
In the presence of these waves, STFA produces a quasi-thermal electron spectrum below $E_t$.  Above $E_t$, the 
scattering mode required to produce a power law spectrum is constrained but undetermined.  This remains a key unresolved issue for 
STFA.  In order to produce a power law, the pitch angle scattering length must be inversely proportionl to the electron
energy $E$ \citep{selkblack}

\begin{equation}
\frac{\lambda_p}{L} = \frac{\Gamma}{E},
\end{equation}
where $\lambda_p$ is the pitch angle scattering length scale, $L$ is the linear size of the acceleration region, and $\Gamma$ 
is a proportionality constant which is fixed by the scattering physics.  The resulting spectrum has the form

\begin{equation}
N(E) \propto E^{-\gamma_{st}},
\end{equation}
where $N(E)$ is the number of electrons with kinetic energy $E$, and the spectral index $\gamma_{st}$ is given by
\begin{equation}
\gamma_{st} = 1+ \frac{\Gamma}{4mv^2_A},
\end{equation}
where $m_p$ is the proton mass and
$v_A$ is the local mean Alfv\'en speed in the plasma.
Variations in $\Gamma$, resulting from local changes in the pitch angle scattering length scale, produce the observed range of 
spectral indexes among flares.  Typically, STFA electron (X-ray) spectra have a mean index of  $4.5(3.5)$ \citep{Bromund}.
 
\subsection{Power law acceleration solely by fast shocks}

	Shock Fermi acceleration had been considered as a means of producing the power law 
electrons required to explain flare X-rays, for example \citet{Blackman97, Tsuneta2}.  A stationary fast shock can form at the 
point where the turbulent reconnection outflow plows into the loop.  This shock, located just above the 
top of the loop, can accelerate electrons into a power law distribution.  

	Shock Fermi acceleration is well studied \citep{Bell, Bell2,  blandford}.  
In the standard theory, charged particles gain energy from repeated transits across the shock.  
Since there is a net energy gain in each cycle, the process is first order and results in a rapid 
acceleration.  The energy spectrum produced by shock Fermi acceleration in impulsive solar flares in the non-relativistic regime 
is a power law of the form 

\begin{equation}
N(E) \propto E^{-\delta_r},
\label{indexshock}
\end{equation}
where $\delta_r=2(r+2)/(r-1)$, and $r$ is the compression ratio across the shock (e.g. Jones \& Ellison 1991).
Furthermore, the shock compression ratio is related to the downflow Mach number $\mathcal{M} \equiv v_f/c_s$

\begin{equation}
r = \frac{(\gamma+1)\mathcal{M}^2}{(\gamma+1+(\gamma-1)(\mathcal{M}^2-1))},
\label{compression}
\end{equation}
where $\gamma$ is the adiabatic index of the plasma.
From simple theoretical estimates, it is expected that the shocks are weakly compressive (Blackman \& Field 1994; Blackman 1997).
To fit the observationally inferred electron spectrum \citep{Brown, Bromund} of $\left<\delta_r\right> = 4.5$
requires a compression ratio $r = 2.6$, significantly less than the maximum of $r=4$ for a non-cooling shock, and consistent 
with theoretical predictions.

Weak fast shocks are expected, but standard shock Fermi acceleration also requires injection of high energy electrons which satisfy 
two conditions: their gyro-radius must be larger than the shock thickness (roughly the thermal proton gyro-radius) to see the shock
as thin, and the electrons must be energetic enough to scatter off of upstream Alfv\'en waves \citep{Bell, blandford}.  The 
threshold energy, $E_{in}$ for injection at impulsive solar flare shocks is 
found to depend on the angle of obliquity between the normal to the shock surface and magnetic field lines, and ranges from $2$keV at 
$85$deg to $10$keV at $0$deg
\citep{Tsuneta2}.  Herein we propose that STFA, in the presence of pitch scattering by whistler waves, 
is the shock acceleration injection mechanism; it accelerates electrons from the background plasma to energies above the proton 
thermal energy.  Note that this scenario differs from that of \citet{Tsuneta2} in which whistler waves are also called upon as a 
scattering mechanism, but the acceleration is single stage; all electron acceleration takes place at the shock in their model.

The electron spectrum from fast shock acceleration  following STFA in the presence of whistler waves is 
can be reproduced in some STFA models without shock acceleration, although the spectrum
from shocks depend only on the compression ratio.
However, in the case with shocks,  the 
transition to a power law spectrum occurs at the shock injection energy, $E_{in}$, instead of at $E_t$.  
A similarity between STFA and shocks is that both
have difficulty in producing downward spectral breaks from an initially steep spectrum
but can produce upward breaks (hardening).  It has been
established 
by \citet{Bell2} for example, that processing of an electron population by a shock can harden an already existing power law 
distribution, with spectral index $\gamma_{in}$, or reduce the index below $\gamma_{in}$, but cannot soften it, or raise the index 
above $\gamma_{in}$. 
If there is a power law spectrum downstream of a shock with a low energy cutoff below the injection energy 
and the shock is sufficiently strong, the upstream spectrum will be harder above $E_{in}$ but remain unchanged below $E_{in}$.  
This property is important to the shock reprocessing model.

\section{The shock reprocessing model}

In the following section, we present the shock reprocessing model and compare it to the trap precipitation model of 
Brown and Melrose (1976).  

\subsection{Overview of the model}

Observations of non-thermal X-ray emission at the footpoints of solar flares indicate that the observed emission
can be divided into two components, a smoothly varying part and a rapidly varying set of pulses \citep{Aschwanden}.  Observations
in multiple energy bands show that for the smooth component, variations in X-ray intensity arrive first at lower energies, and
later at higher energies.  The pulses exhibit an opposite time delay structure; variations are seen first in higher energy
bands.  The series of papers by Aschwanden and collaborators \citep{Aschwanden, A96a, A96b, A97, A98a, A98b, A99}
 models the X-ray arrival time observations in solar 
flares within the trap precipitation model.  We propose the shock reprocessing model as an alternative.  

\begin{figure}
\epsscale{0.9}
\plotone{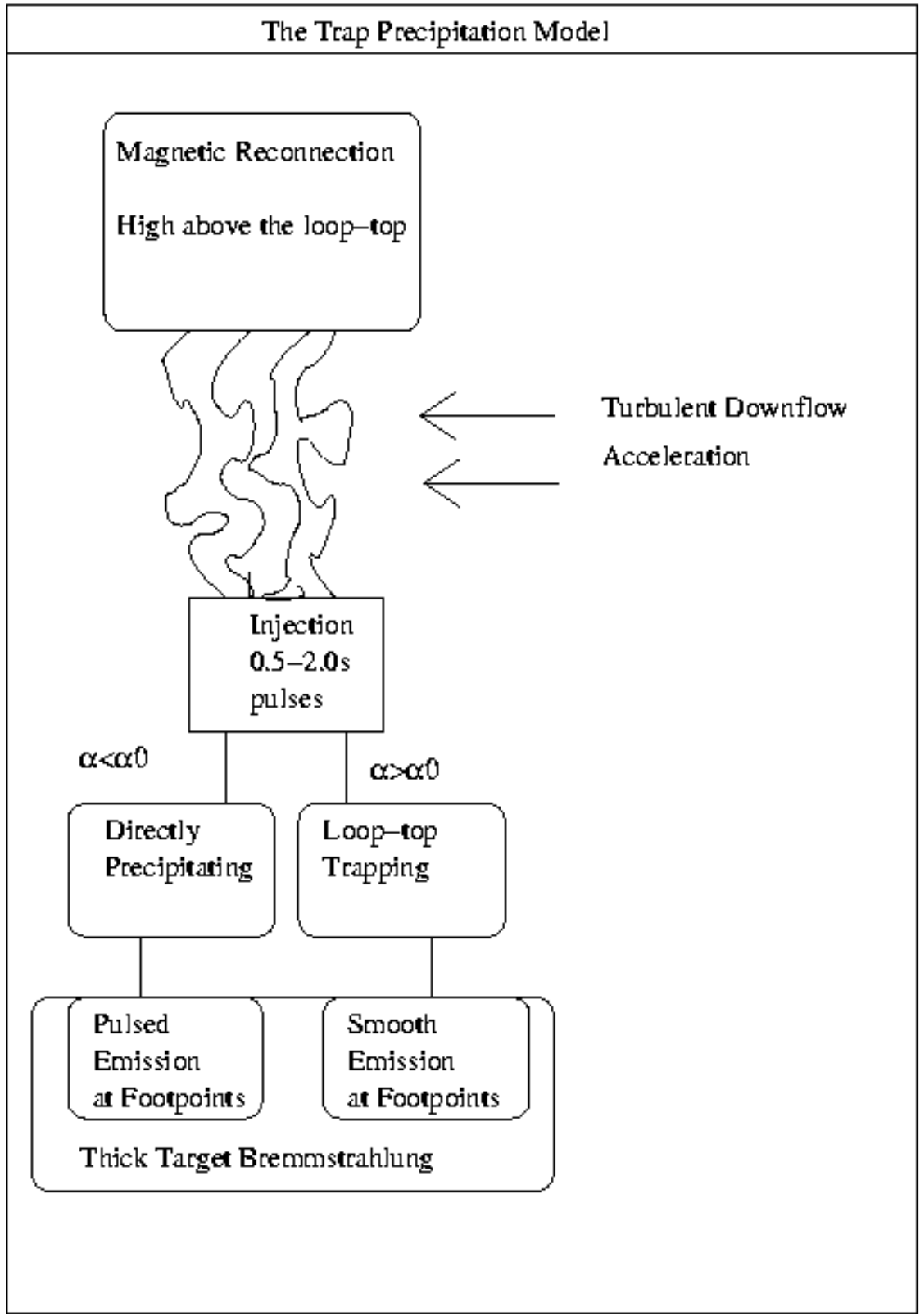}
\caption{A flowchart of the trap precipitation model. $\alpha$ is the electron pitch angle cosine and $\alpha_0$ is the 
critical pitch angle cosine for trapping in the loop.}
\label{trapprecip}
\end{figure}

In the trap precipitation model (Fig. \ref{trapprecip}), reconnection occurs high above the flare loop.  
Electrons are then accelerated
to high energies above the loop over a short time scale, resulting in a pool of non-thermal electrons in a near equilibrium 
distribution.  
An injection event somehow loads a pulse of electrons, with an energy 
spectrum matching that of the pool above the loop-top, onto the loop.  Injection occurs over a very short time, $50$ms, matching the 
observed X-ray pulse durations.  Injected electrons travel down the flare loop, where they encounter a magnetic mirror as the loop 
field lines converge.  Electrons with sufficiently high pitch angle relative to the local field are trapped at the mirror, while 
those with small pitch angle stream through.  This divides the electrons into two populations: directly precipitating, and trap 
precipitating electrons.  The directly precipitating electrons do not reflect at the mirror, and proceed directly to the dense
footpoint region.  There they emit X-rays via thick target bremsstrahlung; these electrons retain the short pulse time scale of the 
injection, and thus produce the pulsed emission component.  Since the pulse electrons are injected simultaneously at all energies, 
the main
time dispersion of a pulse results from time of flight down the loop, with higher energy (faster) electrons arriving earlier than 
lower energy (slower) electrons, consistent with the pulse observations.  The second population, trap precipitated electrons, are mirrored
at the magnetic field compressions due to their large pitch angles.  These electrons remain trapped between the mirrors on opposite 
sides of the flare loop until such time as their pitch angles scatter to sufficiently small values that they can pass through the 
mirrors.  \citet{A97} find that the slow time delays are consistent with loop-top trapping, where escape from the trap is governed
by the collisional scattering time which varies as $E^{3/2}$. 
Because the trapping time is significantly longer than the injection time, electrons from many injections escape 
together, creating a smooth component.  As a result of the positive energy dependence of the trapping time, lower energy electrons 
escape the trap earlier than higher energy electrons, and the smooth emission component exhibits variations at lower energies first.
The trap precipitation model offers an explanation for the two component emission, as well as the time delay data. However, it is 
not unique. 

\begin{figure}
\epsscale{0.95}
\plotone{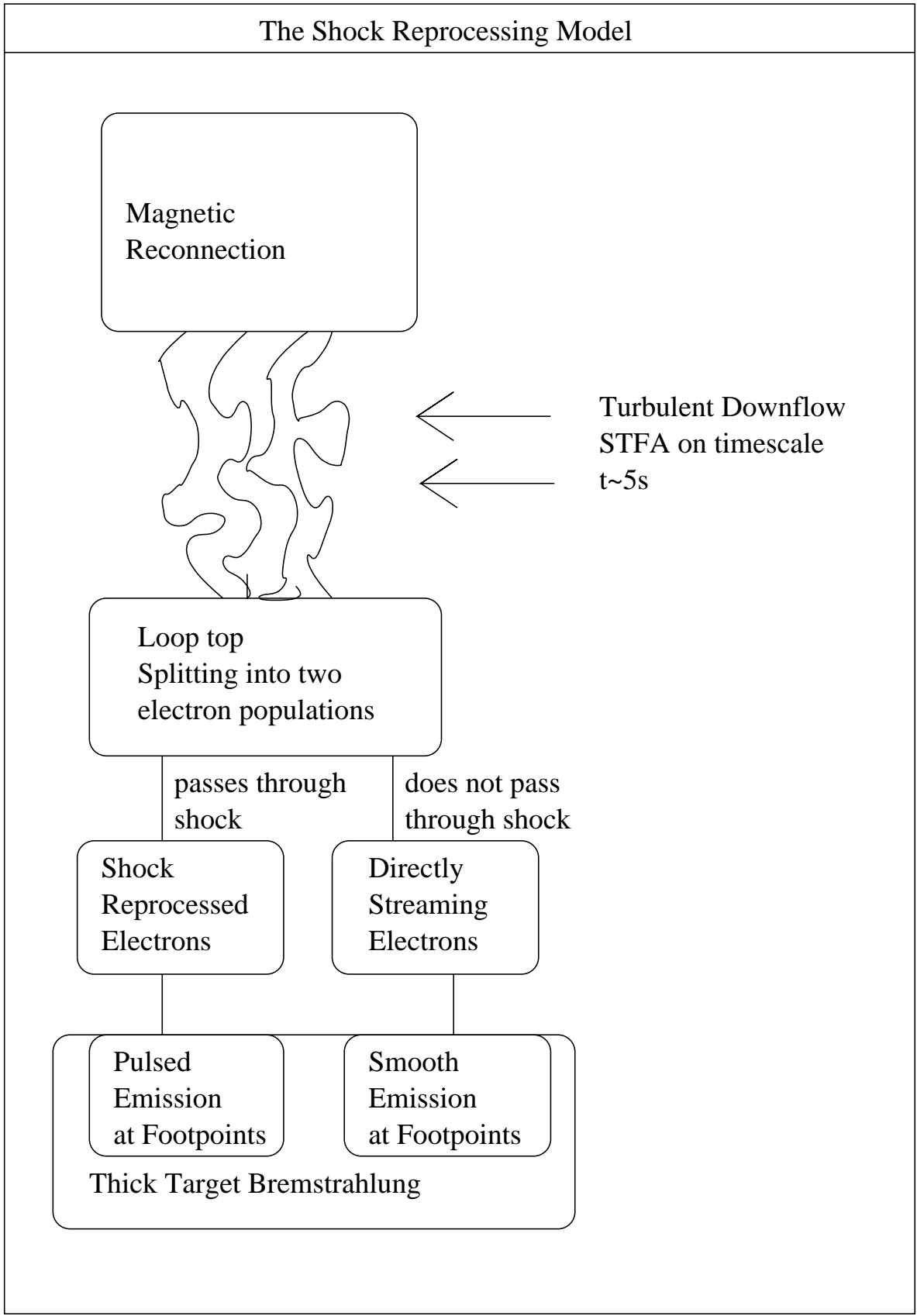}
\caption{A flowchart of the shock reprocessing model.}
\label{reprocessing}
\end{figure}

The shock reprocessing model is an alternative to trap precipitation.  An outline of the model is shown in Fig. \ref{reprocessing}.
As in trap precipitation, the shock reprocessing model begins with reconnection high above the flare loop, 
followed by acceleration above the loop-top.  However, in the shock reprocessing model the initial acceleration is slow and 
consistent with STFA in the reconnection outflow, occurring on a time scale of $~5$s \citep{chandran, selkblack}.  
Following STFA, the electrons flow into the region just above the loop-top.  At or just above the loop-top a weakly compressive 
fast shock forms \citep{Blackman94, Tsuneta}.  The shock can vary in spatial extent and compression ratio.  A fraction of the STFA
accelerated electrons also pass through the shock and are accelerated a second time (reprocessed).  The remainder pass directly to 
the flare footpoints. 

Like the trap precipitation model, the shock reprocessing model involves two populations of electrons, but in this case they are the 
directly streaming (STFA only) electrons and the shock reprocessed electrons.  The directly streaming electrons produce the 
smooth X-ray component.  Since there is no trapping or injection within this model, 
variability in the produced emission results solely from the acceleration process. 
The $\sim 5$ s acceleration time for STFA \citep{selkblack, chandran} is consistent with the delay time observed between 
$25$keV and $150$keV electrons in flare observations \citep{Aschwanden, A96a, A96b, A97, A98a, A98b, A99}.   
The reprocessed electrons are those which pass through the shock and undergo a second, rapid acceleration. 
Subsequently, they travel without coronal trapping to the footpoints.  Because the shock acceleration time scale is extremely short,
reprocessed electrons can be considered effectively injected at the loop-top simultaneously across all energies.
The number of electrons which are reprocessed is determined by the spatial extent of the shock.  We proceed by 
developing a phenomenological procedure for determining pulse strengths, discussing shock formation and geometry as further 
constraints on the model, and predicting the location of the coronal emission site coincident with the STFA and cooling in a region 
above the loop-top.  

The absence of an above the loop-top, or loop-top trap is a key distinction between shock reprocessing and trap precipitation models.  
Expansion of the closed loop 
is often assumed to occur near the loop-top and is a requirement of the trap precipitation model
to create the trap in the form of a magnetic mirror.
Recently however, \citet{Bellan1} presents observational and theoretical evidence against the formation of a
trap, finding instead that the loops are very nearly axially symmetric
from footpoint to apex.  This poses a challenge to the static trap
models which require  narrowing loop cross sections 
at the footpoints and broader loop cross sections at the apex.

\subsection{Parameterizing the model}

The filling fraction of the shock, $F$, can be understood as both the fraction of the downflow 
cross section which encounters the shock, and 
as the fraction of non-thermal electrons which are reprocessed.  Note that this sets the firm upper bound $F=1$.
The physics relating $F$ to shock formation is deferred to section 
3.4.  Here, we constrain $F$ from observations and power law models of the electron and x-ray spectra.  
We begin by assuming that the electron spectrum resulting from STFA is a single power law, given by

\begin{equation}
N_s = {N_{0s}} \left( \frac{E}{E_{0s}} \right)^{-\delta_s},
\label{powerlawSTFA}
\end{equation}
where $N_s$ is the total number of electrons per unit energy per unit time passing through a slab of unit surface area coplanar 
with the shock (henceforth we refer to similar quantities as areal fluxes), and is therefore the total non-thermal electron population, 
as a function of energy.  $N_{0s}$ is the areal flux of electrons at energy $E_{0s}$, the lower threshold energy for the onset 
of the STFA power law spectrum, and $\delta_s$ is the spectral index of the STFA electrons.  The total number of non-thermal electrons 
is then given by

\begin{equation}
N_T = A \int_{E_{0s}}^{\infty} N_s dE = \frac{A E_{0s} N_{0s}}{\delta_s - 1},
\label{NT}
\end{equation}
where $N_T$ is the total number of non-thermal electrons and $A$ is the area of the downflow.  Because of the injection criterion, 
shock reprocessing does not increase the total number of non-thermal electrons.  As a result, the total number of reprocessed 
non-thermal electrons, $N_{p}$, can be written as 

\begin{equation}
N_{p} = F A \int_{E_{0r}}^{\infty} N_s dE = F \frac{A E_{0r} N_{0s}}{\delta_s -1} \left(\frac{E_{0r}}{E_{0s}}\right)^{-(\delta_s -1)} 
                                          = F  E_{0r} \left(\frac{E_{0r}}{E_{0s}}\right)^{-(\delta_s -1)} N_T,
\label{NP1}
\end {equation}
where we have taken into account only those electrons with energy above the shock injection threshold, $E_{0r}$, and assume
$E_{0r} > E_{0s}$.  

We can obtain a second expression for $N_p$ by
assuming a power law of similar form to Eq. (\ref{powerlawSTFA}) for the reprocessed electrons,

\begin{equation}
N_r = N_{0r} \left( \frac{E}{E_{0s}} \right)^{-\delta_r},
\label{NR}
\end{equation}
where quantities with the subscript $r$ are defined similarly to those with an $s$ in Eq. (\ref{powerlawSTFA}), 
but for the shock reprocessed population.
The total number of electrons in the reprocessed population, $N_p$, is obtained analogously to Eq. (\ref{NT}) by integrating 
over energy 

\begin{equation}
N_{p} =  FA \int_{E_{0r}}^{\infty} N_r dE
      =  F \frac{A E_{0r} N_{0r}}{\delta_r - 1}.
\label{NP2}
\end{equation}
Setting Eqs. (\ref{NP1}) and (\ref{NP2}) equal yields

\begin{equation}
N_{0r} = N_{0s} \frac{\delta_r -1 }{\delta_s -1}  \left(\frac{E_{0r}}{E_{0s}}\right)^{-(\delta_s -1)}.
\label{N0related}
\end{equation}
For thick target bremsstrahlung at the footpoints, the photon spectrum is likewise a power law

\begin{equation}
M_x = M_{0x}  \left( \frac{E}{E_{0x}} \right)^{-\alpha_x},
\label{Meqn}
\end{equation}
where $x$ can be either $r$ or $s$ for the shock and STFA components, $M_x$ is the areal photon flux at energy $E$, and $M_{0x}$ 
is the areal photon flux at energy $E_{0x}$.  The spectral index
for the photons is related to that of the electrons by $\alpha_x = \delta_x - 1$ \citep{Brown}.  
Combining this with Eqs. (\ref{powerlawSTFA}) and (\ref{NR}) yields the additional constraint

\begin{equation}
\left(\delta_s -1\right)\frac{M_{0s}}{N_{0s}} = \left(\delta_r - 1 \right)  \frac{M_{0r}}{N_{0r}}.
\label{M0related}
\end{equation}
Integrating both the STFA and reprocessed spectra of Eq. (\ref{Meqn})
from the minimum observed energy $E_{min}$, where we assume $E_{min} > E_{0r} > E_{0s}$, gives the 
areal photon flux for both the STFA and reprocessed components. 

\begin{equation}
M_T = A \int_{E_{min}}^{\infty} M_s dE = A \frac{M_{0s} E_{0s}}{\delta_s -2} \left(\frac{E_{min}}{E_{0s}}\right)^{-(\delta_s -2)},
\label{M_T}
\end{equation}
\begin{equation}
M_p = A \int_{E_{min}}^{\infty} M_r dE = A \frac{M_{0r} E_{0r}}{\delta_r -2} \left(\frac{E_{min}}{E_{0r}}\right)^{-(\delta_r -2)},
\label{M_p}
\end{equation}
where $M_T$ and $M_p$ are the energy integrated areal photon flux from the STFA and shock reprocessed populations respectively.  
Using Eqs. (\ref{N0related}), (\ref{M0related}), (\ref{M_T}), and (\ref{M_p}) we can rewrite $M_p$ in terms of $M_T$

\begin{equation}
M_p = \left(\frac{\delta_s -2}{\delta_r -2}\right) E_{0r}^{\delta_r - \delta_s} E_{min}^{\delta_s-\delta_r} M_T.
\label{MpofMt}
\end{equation}
 
Given the above, we can construct photon fluxes for the pulsed and smooth X-ray components given values of the free parameters of 
the model: $F$, $\delta_r$, and $\delta_s$.  We define the smooth and
pulsed emission as follows.  The smooth emission is the total rate of photon counts at the footpoints in the absence of pulses, which
is simply given by Eq (\ref{M_T}) times the area of the downflow.  
The pulse emission, $M_P$, is the enhancement above the smooth emission, which we take at the peak of the pulse, and is given by 

\begin{equation}
M_P = F  M_p - F M_T = 
 F \left( \left(\frac{\delta_s -2}{\delta_r -2}\right) E_{0r}^{\delta_r - \delta_s} E_{min}^{\delta_s-\delta_r} - 1 \right) M_T,
\label{PulseFormula}
\end{equation}
which is the total emission by the reprocessed electrons, $F M_p$, less the emission that would have been produced by those electrons
if they were not reprocessed, $F M_T$.   
For further simplicity, we choose to measure energy in units of $E_{0r}$.  Thus, we can write

\begin{equation}
M = \frac{M_P}{M_T} = F \left(\left(\frac{\delta_s -2}{\delta_r -2}\right) E_{min}^{\delta_s-\delta_r} - 1 \right).
\label{Mpeqn}
\end{equation}
The model produces pulses of magnitude $M = M_P/M_T$ determined by four parameters: the shock filling fraction $F$, the STFA 
electron spectral index $\delta_s$,  the reprocessed electron spectral index $\delta_r$, and the minimum observed energy $E_{min}$.

\subsection{Fitting the observed pulse strengths}

\begin{figure}
\epsscale{1.0}
\plotone{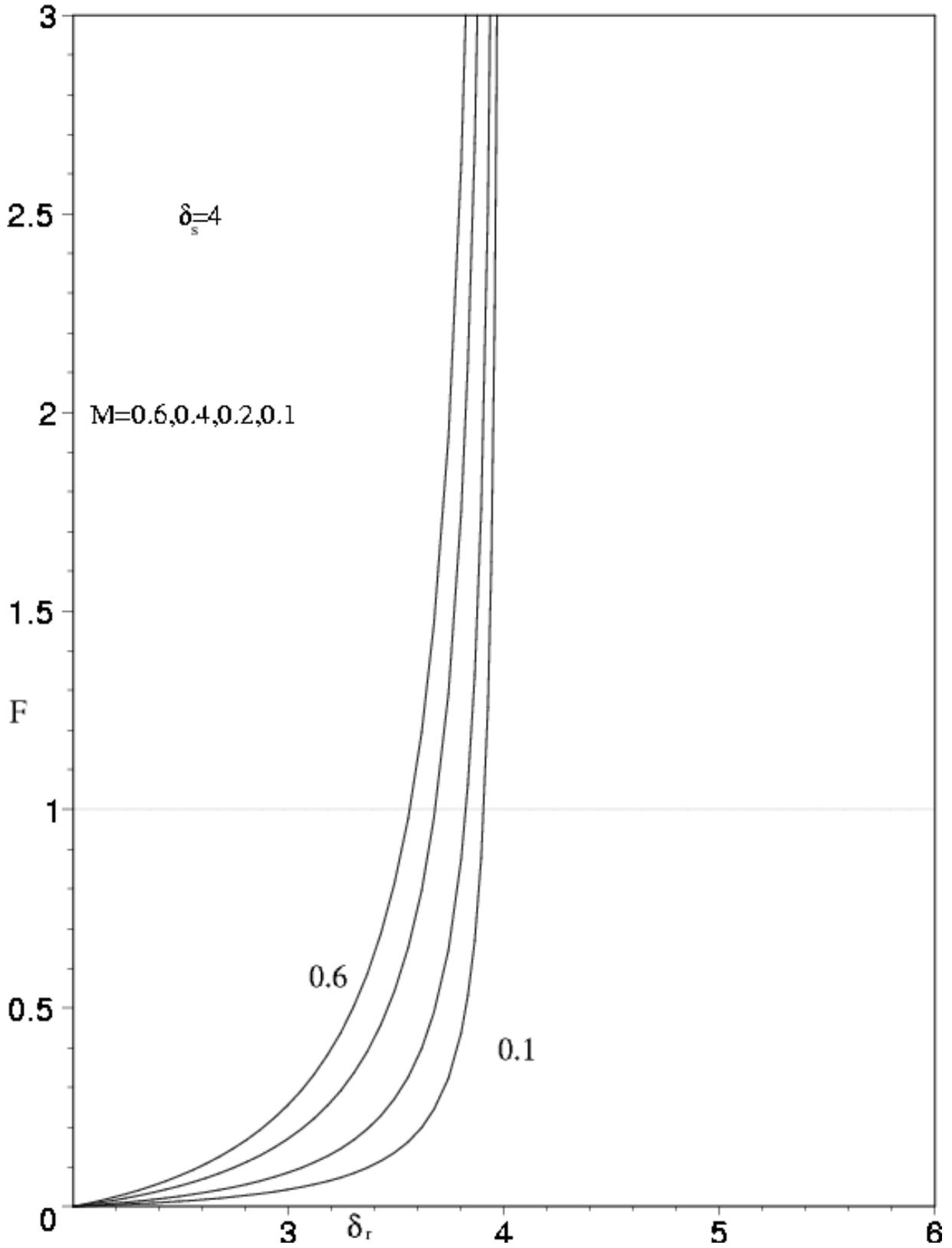}
\caption{Curves of constant $M$.  From top to bottom: $M=0.6, 0.4, 0.2, 0.1$.  $E_{min} = 5/3$.  $\delta_s = 4$
Only $F \leq 1$ corresponds to physically realizable states.}
\label{Mcurvesbase}
\end{figure}

Fig. \ref{Mcurvesbase} shows a set of curves of constant pulse strength ($M = 0.1, 0.2, 0.4, 0.6$) in $\delta_r$ vs $F$ space.  We
have taken $E_{min} = 5 E_{0r}/3$ and $\delta_s = 4$, consistent with the low energy observational cutoff of $E=25keV$ in 
\citet{Aschwanden} and a presumed non-thermal cutoff energy of $\backsim 15keV$.  
A number of important properties of the model are evident.  First, the curves diverge 
asymptotically toward $F = \infty$ as $\delta_r$ approaches $\delta_s$.  This is expected from both Eq. (\ref{Mpeqn}) and the 
physics of shock acceleration.  The result of \citet{Bell2} that a shock can harden, but not soften, an electron spectrum indicates that
the shock reprocessing model can only produce pulses via reacceleration at the shock if $\delta_r < \delta_s$.  Since $\delta_r$ is
determined by the shock compression ratio, pulsing (as opposed to steady enhancement) of the reprocessed component requires variation
of the shock compression ratio, or equivalently the downflow Mach number, on the pulse time scale.  A second property of the curves is
that there is an upper limit on $\delta_r$ for any given observed $M$, which is set by the $F=1$ line.  By definition, $F>1$ represents
an unphysical solution, where the number of reprocessed electrons exceeds the total number of non-thermal electrons.  The shock
reprocessing model thus can constrain parameter space.  Any choice of $M$ and $\delta_r$ which predicts $F>1$ represents a physically
unrealizable state.  For example, Fig. \ref{Mcurvesbase} shows that pulses stronger than $M=0.6$ cannot be formed by shocks 
with $\delta_r > 3.4$, while pulses of strength $M=0.4$ can be formed out to $\delta_r=~3.6$.  Generally, the stronger the pulse, the 
harder the limiting shock spectrum.  Finally, notice that all of the curves converge
toward $F=0$ at $\delta_r = 2$.  This results from the increasing difference between the reprocessed and STFA spectra as the shock 
approaches the limiting case of $\delta_r=2$ at a strong shock.  At all spectral indexes, there is a high end cutoff photon energy set
by the total energy budget of the flare.  However, for all values of $\delta_r > 2$, the integrals of Eqs. (\ref{M_T}) and
(\ref{M_p}) are dominated by the lower energy bound, so ignoring the cutoff is acceptable.

\begin{figure}[hbtp]
  \vspace{9pt}
  \centerline{\hbox{ \hspace{0.0in} 
    \epsscale{0.4}
    \plotone{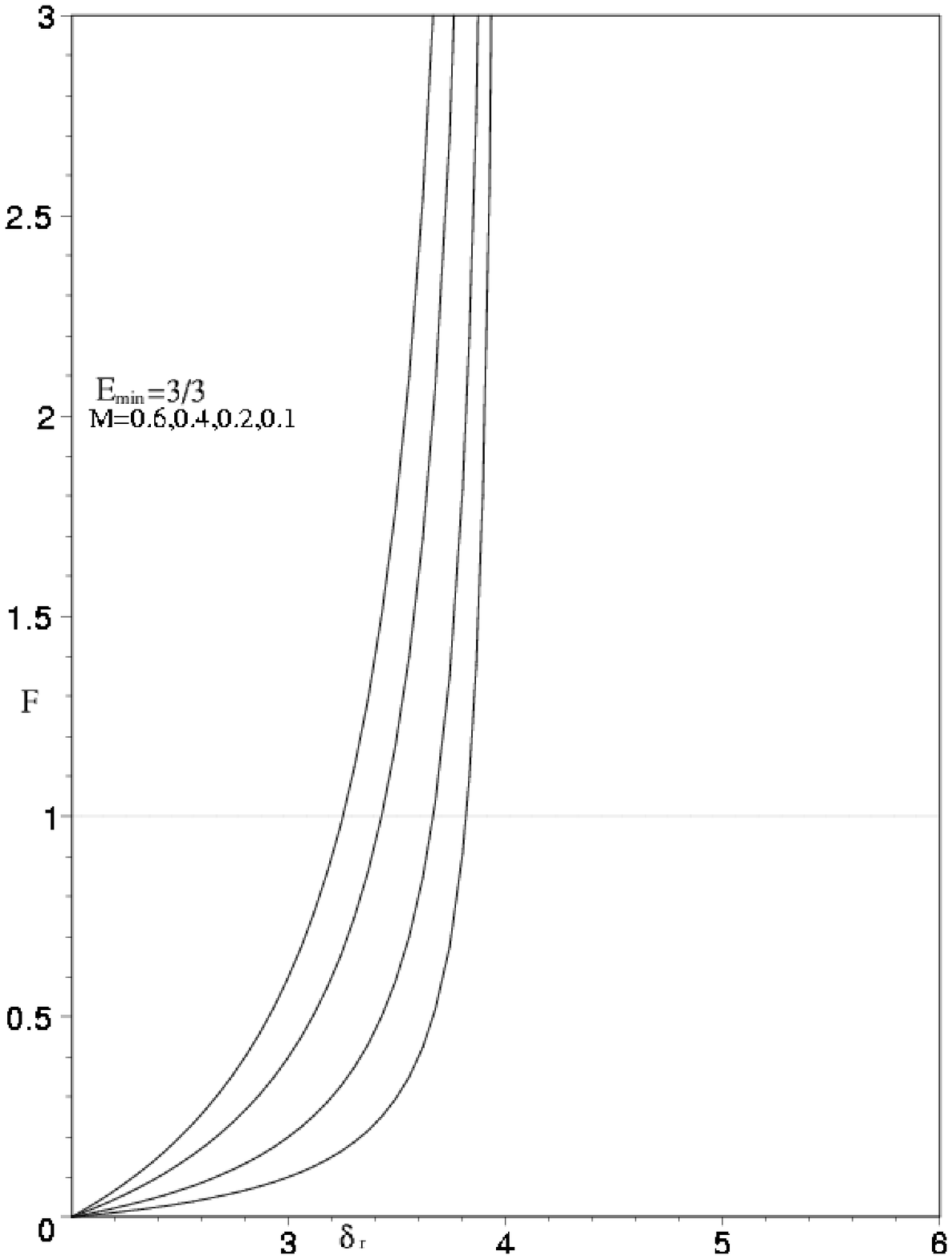}
    \hspace{0.05in}
    \epsscale{0.4}
    \plotone{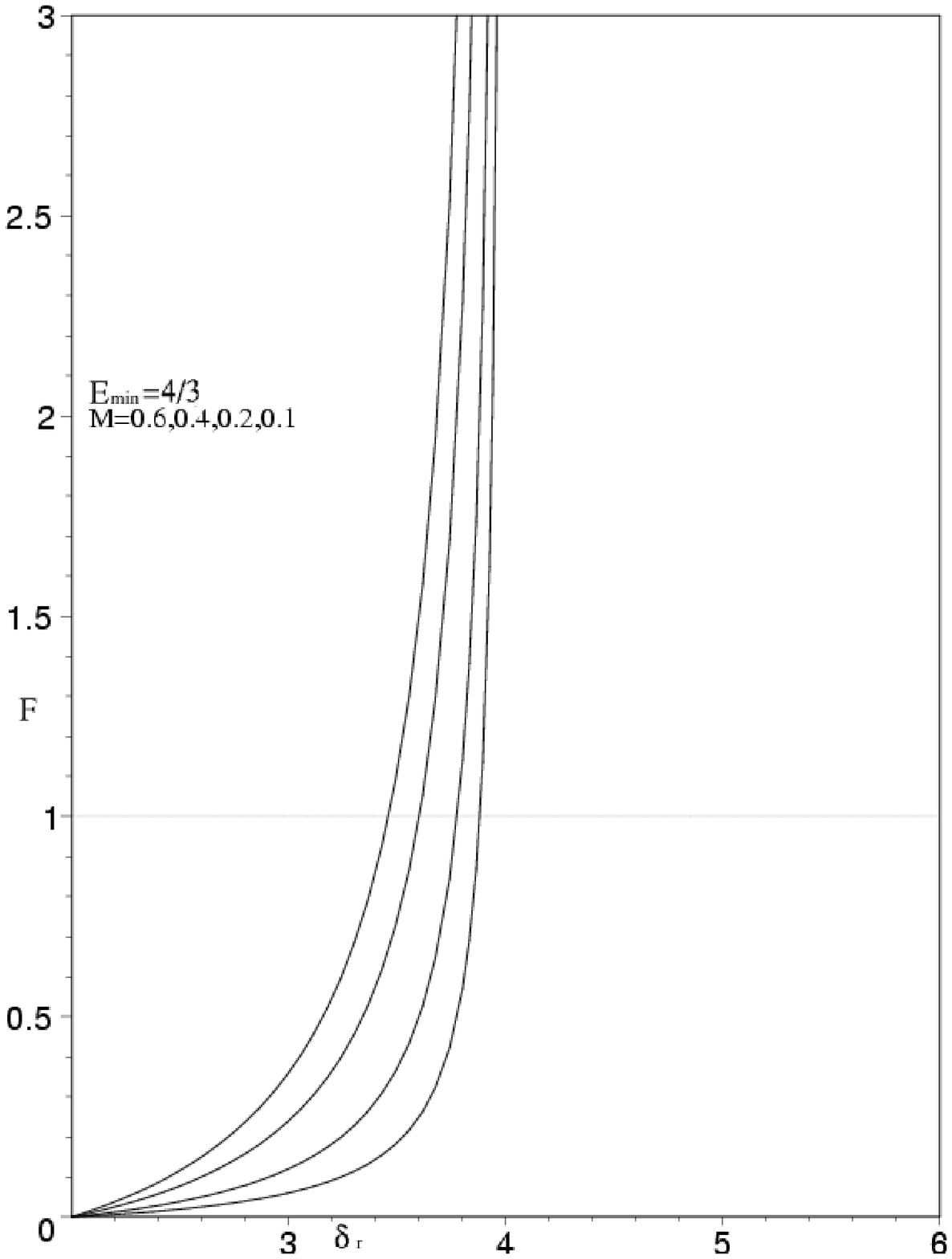}
    }
  }

  \vspace{9pt}
  \hbox{\hspace{1.55in} (a) \hspace{2.60in} (b)} 
  \vspace{9pt}

  \centerline{\hbox{ \hspace{0.00in}
    \epsscale{0.4}
    \plotone{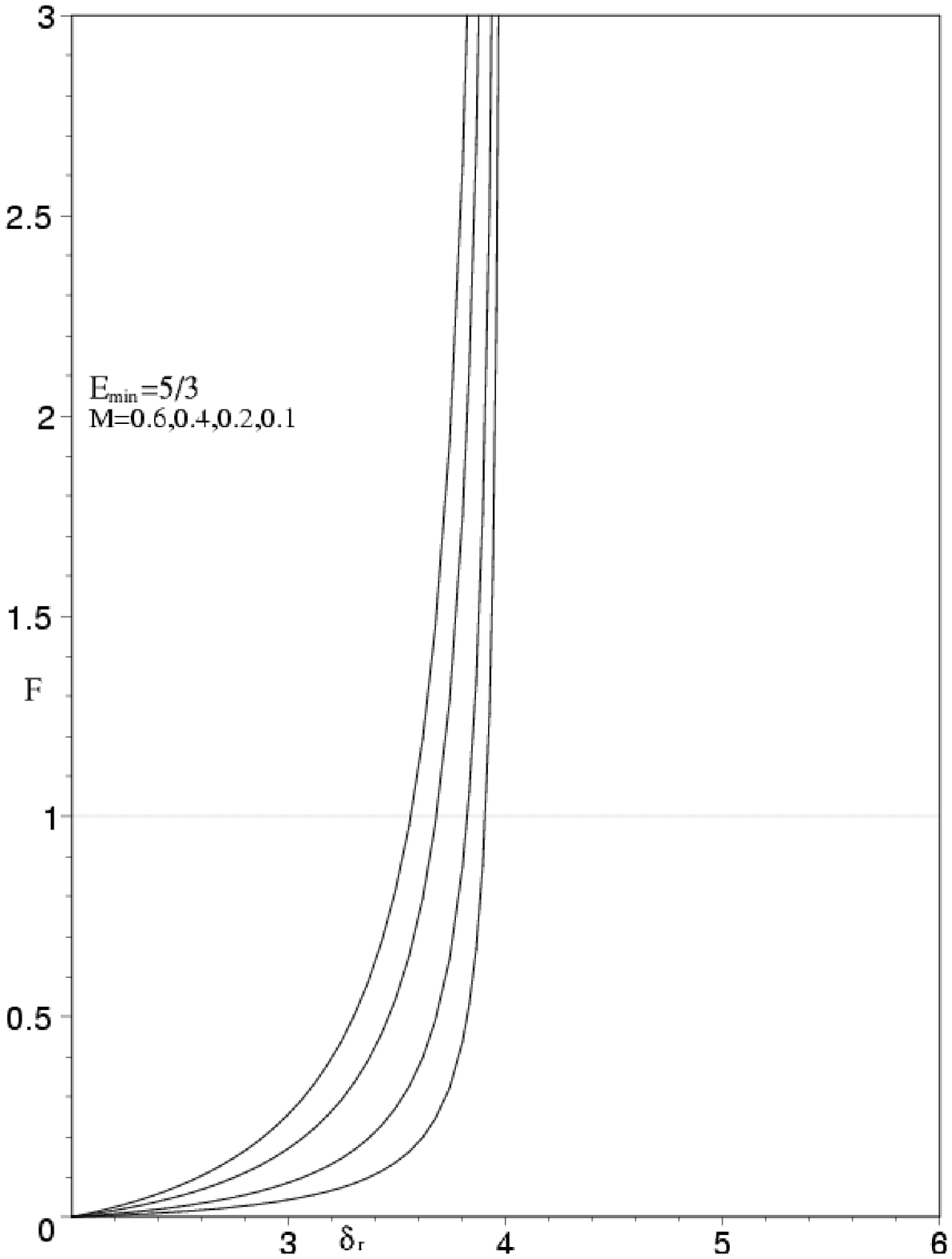}
    \hspace{0.05in}
    \epsscale{0.4}
    \plotone{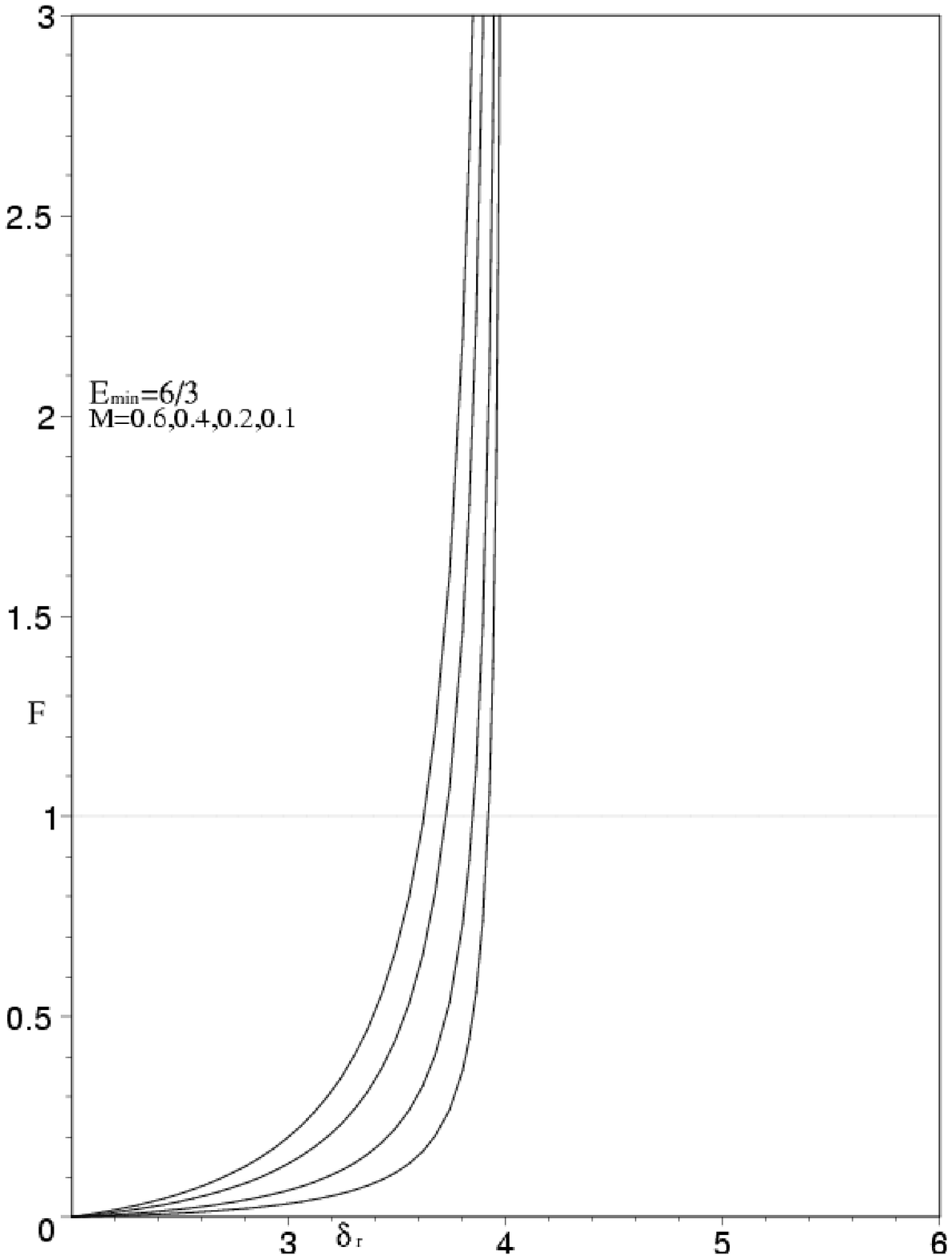}
    }
  }

  \vspace{9pt}
  \hbox{\hspace{1.55in} (c) \hspace{2.60in} (d)} 
\caption{Curves of constant $M$ in $\delta_r - F$ space as $E_{min}$ is varied; $\delta_s =4$. (a) $E_{min} = 3/3$. 
(b) $E_{min} = 4/3$,(c) $E_{min} = 5/3$. (d) $E_{min} = 6/3$.  $E_{min}$ is normalized to $15$keV.  In each panel, curves from left to
right are $M=0.6, 0.4, 0.2, 0.1$.  Only $F \leq 1$ corresponds to physically realizable states.  Increasing $E_{min}$ results in a 
lower required $F$ for a given $M$}.
\label{Evariations}
\end{figure}

\begin{figure}[hbtp]
  \vspace{9pt}

  \centerline{\hbox{ \hspace{0.0in} 
    \epsscale{0.4}
    \plotone{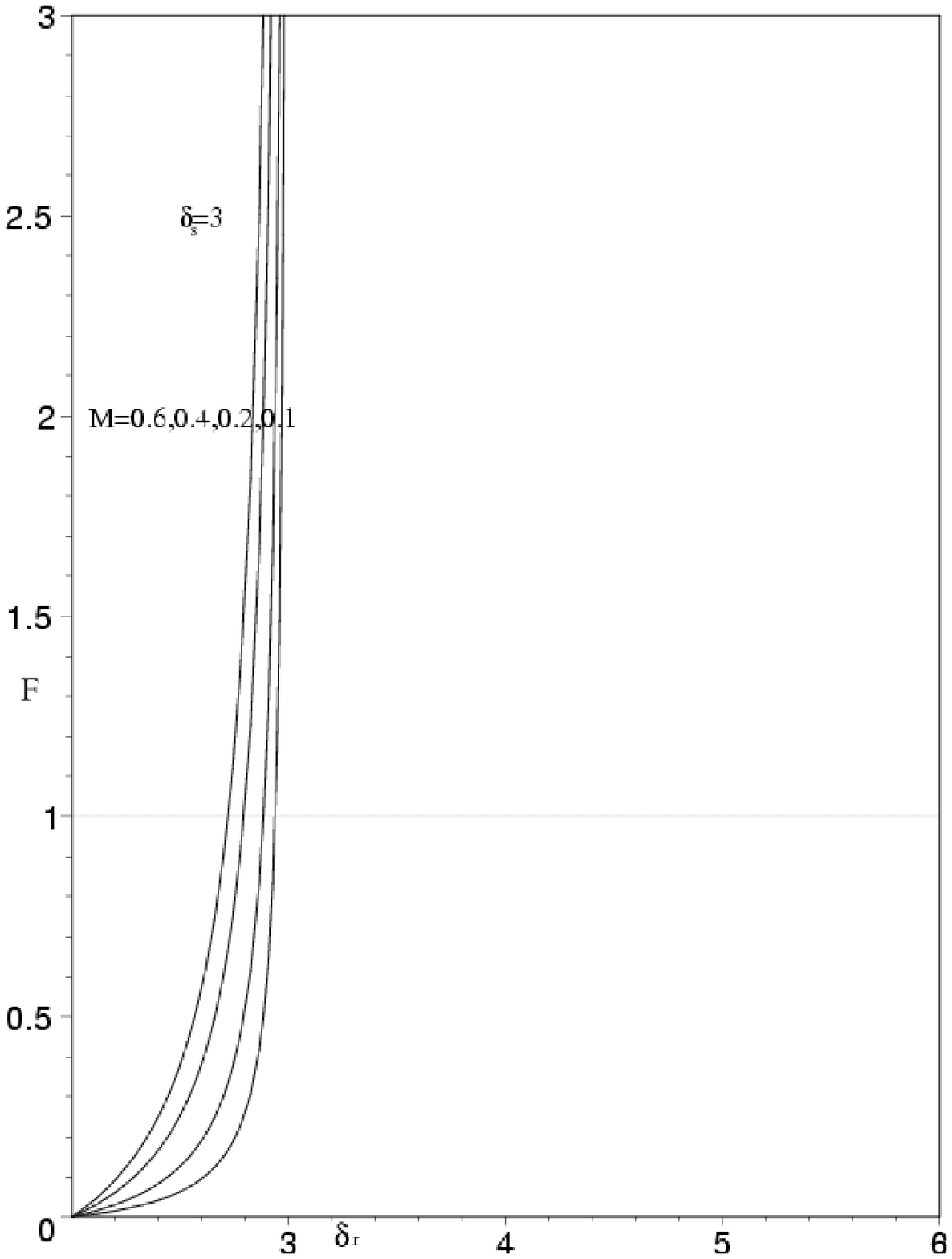}
    \hspace{0.05in}
    \epsscale{0.4}
    \plotone{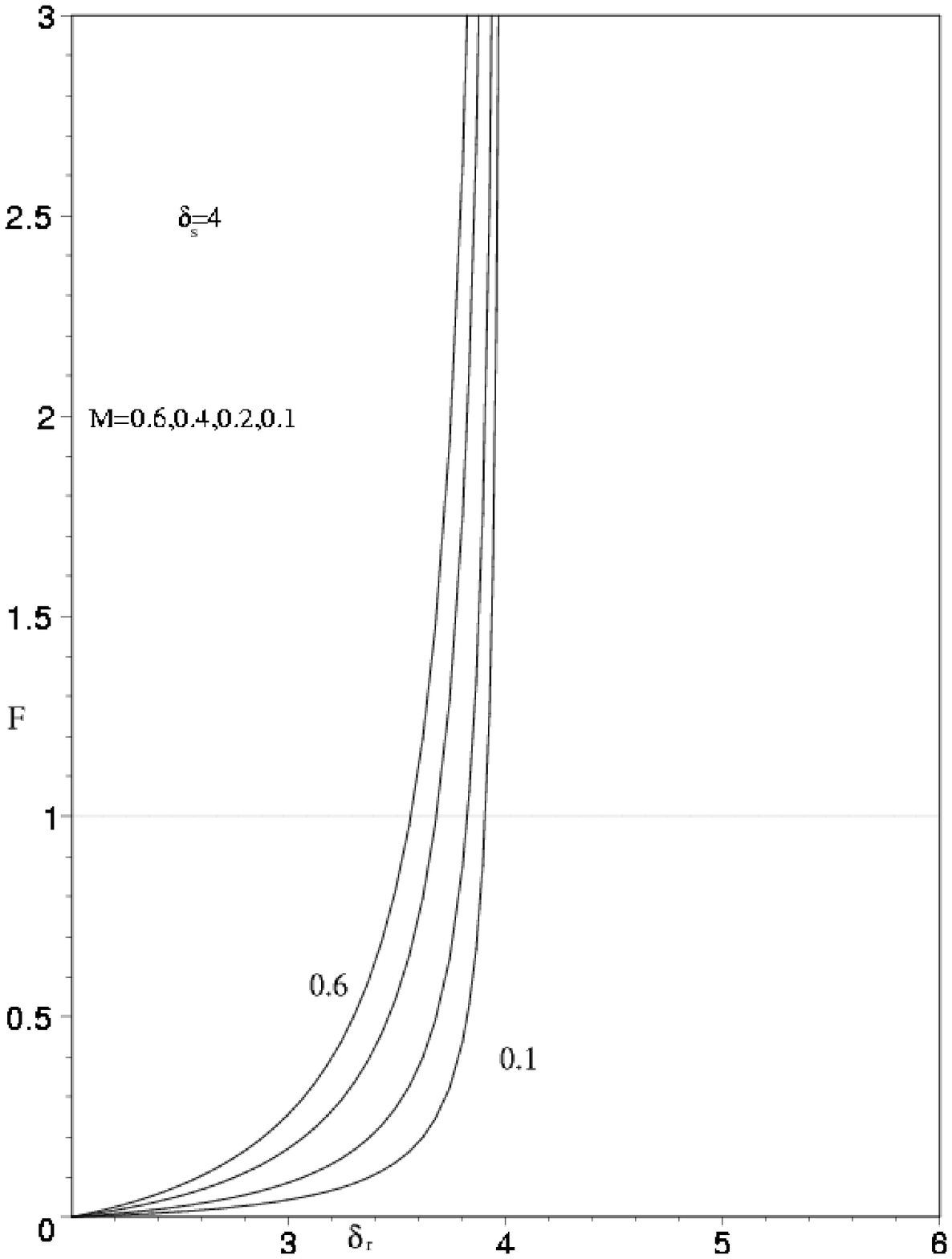}
    }
  }

  \vspace{9pt}
  \hbox{\hspace{1.55in} (a) \hspace{2.60in} (b)} 
  \vspace{9pt}

  \centerline{\hbox{ \hspace{0.00in}
    \epsscale{0.4}
    \plotone{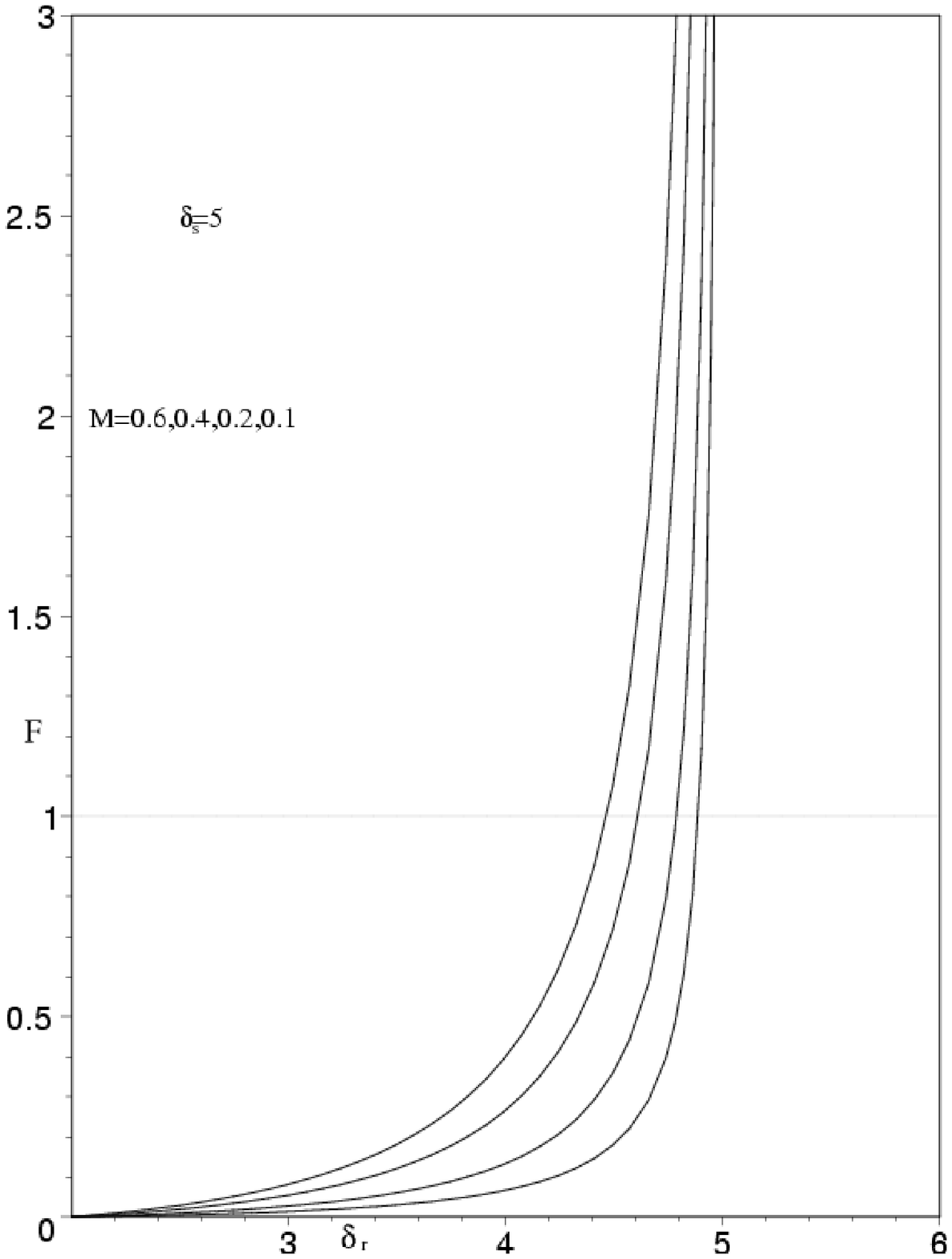}
    \hspace{0.05in}
    \epsscale{0.4}
    \plotone{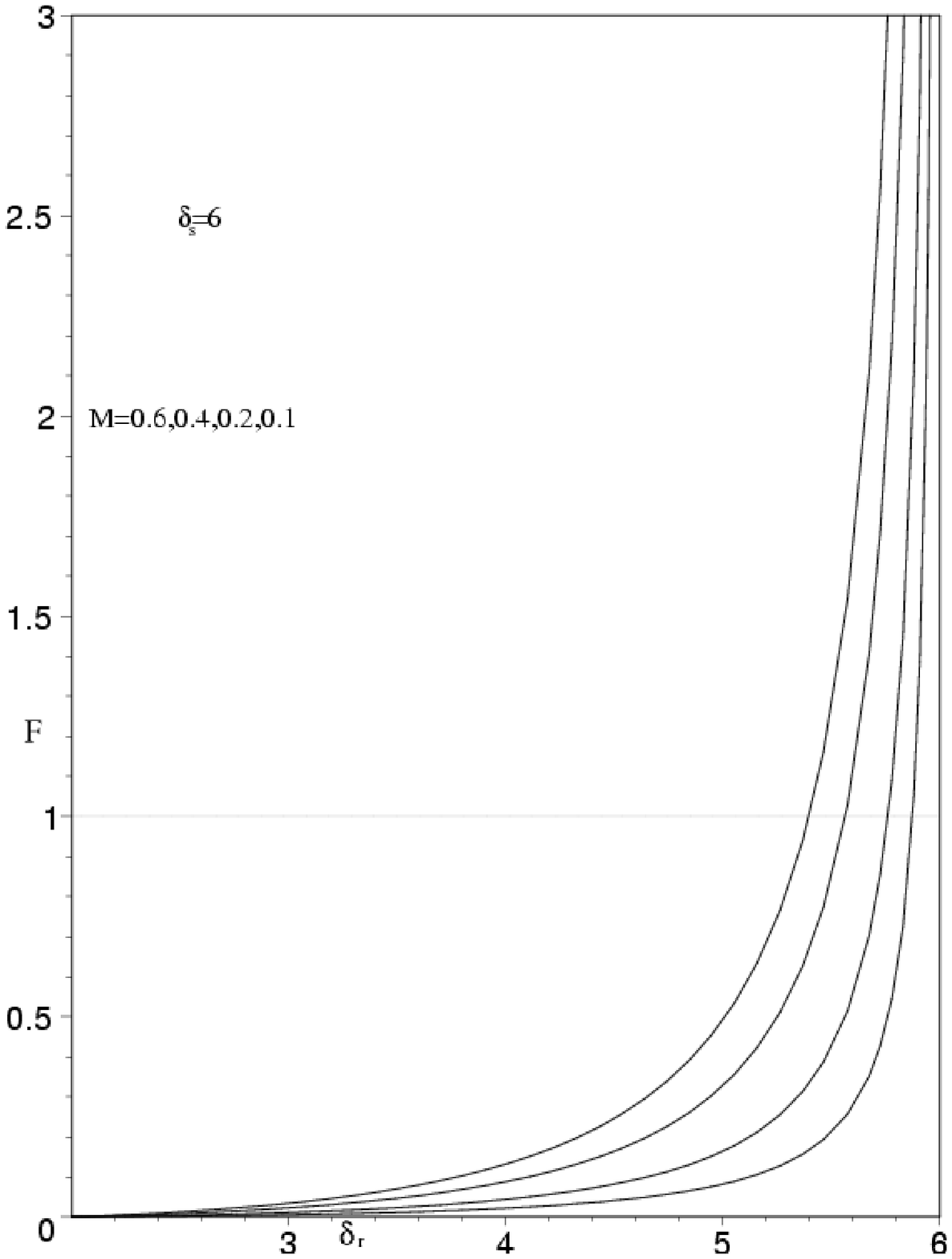}
    }
  }

  \vspace{9pt}
  \hbox{\hspace{1.55in} (c) \hspace{2.60in} (d)} 
\caption{Curves of constant $M$ in $\delta_s - F$ space as $\delta_s$ is varied. (a) $\delta_s = 3$. (b) $\delta_s = 4$,  
        (c) $\delta_s = 5$. (d) $\delta_s = 6$.  $E_{min} = 25$keV in all four panels.  In each panel, curves from left to
right are $M=0.6, 0.4, 0.2, 0.1$.  Only $F \leq 1$ corresponds to physically realizable states.  Increasing $\delta_s$ implies a 
higher $F$ is required to produce a given $M$.}
\label{Svariations}
\end{figure}

Fig. \ref{Evariations} shows the effects of varying the value of $E_{min}$ on the curves of constant $M$ in $F$ and $\delta_r$ 
space.  We have taken $\delta_s = 4$ for each of these plots.  
As we raise $E_{min}$, the curves shift toward smaller values of $F$ for any given $\delta_r$.   The shift results from moving the
observational cutoff energy further from the onset of the reprocessing power law.  The difference between the two 
power law spectra, $M_r$ and $M_s$ is greater at higher energies.  Moving the lowest energy of observation further away from the 
shock injection energy results in stronger pulses.  Therefore, the model predicts that observed pulses will be stronger in higher 
energy bands for any given flare.  We explore the ramifications of this prediction in section 4.
Likewise, Fig. \ref{Svariations} shows the effects of varying $\delta_s$ while keeping $E_{min}$ fixed.  In this set of plots, 
$E_{min} = 5E_{0r}/3$, and $\delta_s$ takes the values $3.5$,$4$, $4.5$, and $5$.  The asymptotic divergence of $F$ always occurs at 
$\delta_r = \delta_s$.  Increasing $\delta_s$ results in the formation of stronger pulses at any given values of $\delta_r$ and $F$.  
  
\subsection{Constraints of shock formation}

\begin{figure}
\epsscale{1.0}
\plotone{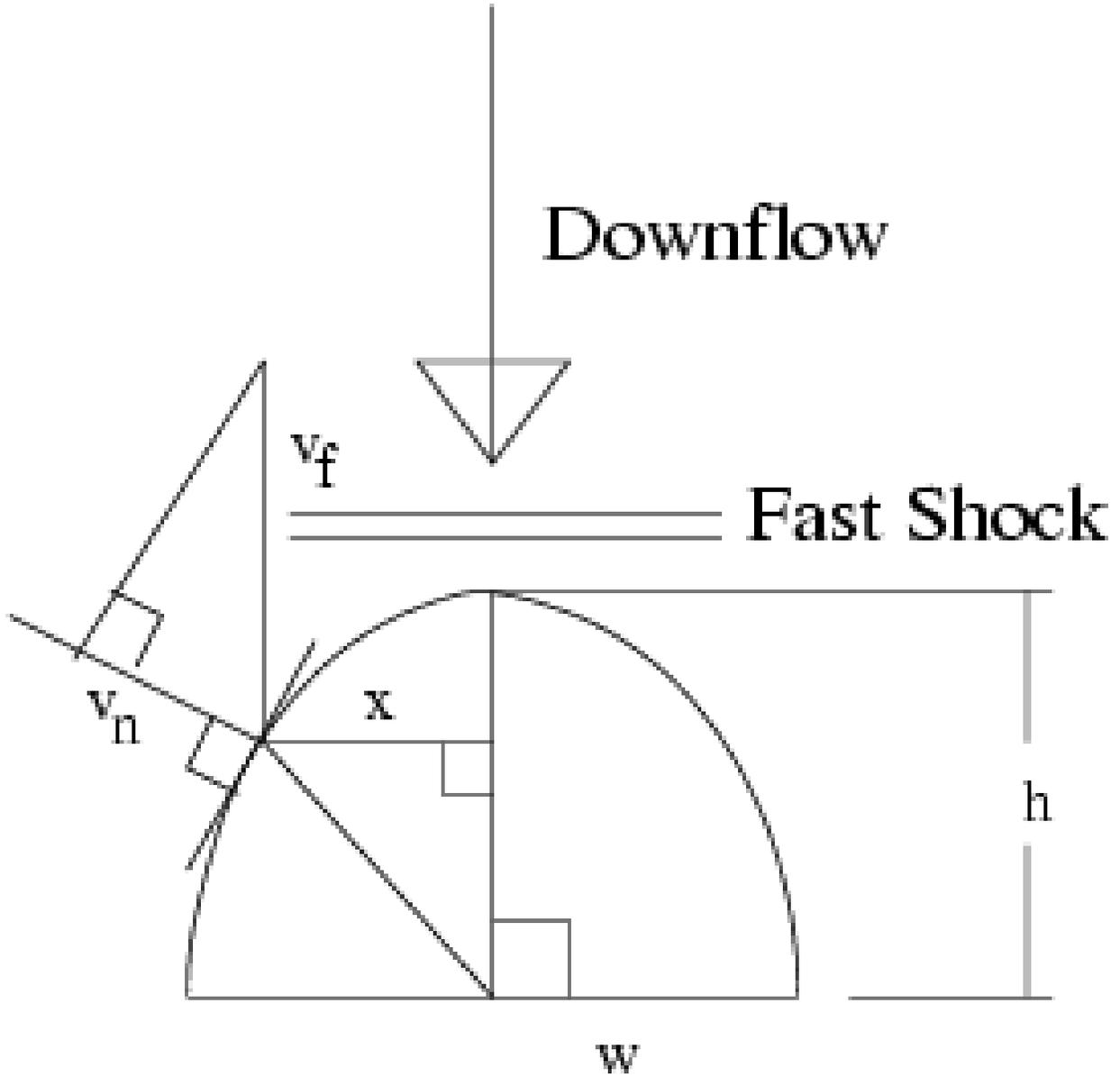}
\caption{A schematic of the shock formation model.  
$v_f$ is the downflow speed, $v_n$ is the component of the speed normal to the
loop, $s$ is the slope of the loop, $w$ and $h$ the loop width and height, and $x$ the distance of the point at which $s$ is evaluated
from the mid-line of the loop.  A fast shock forms above the loop-top for all values of $x$ where $v_n $ is super-fast-magnetosonic. 
In the reconnection downflow, the fast-magnetosonic speed is effectively equal to the sound speed and we can write the 
shock formation condition as $v_n > c_f$.} 
\label{shockform}
\end{figure}

\begin{figure}
\epsscale{1.0}
\plotone{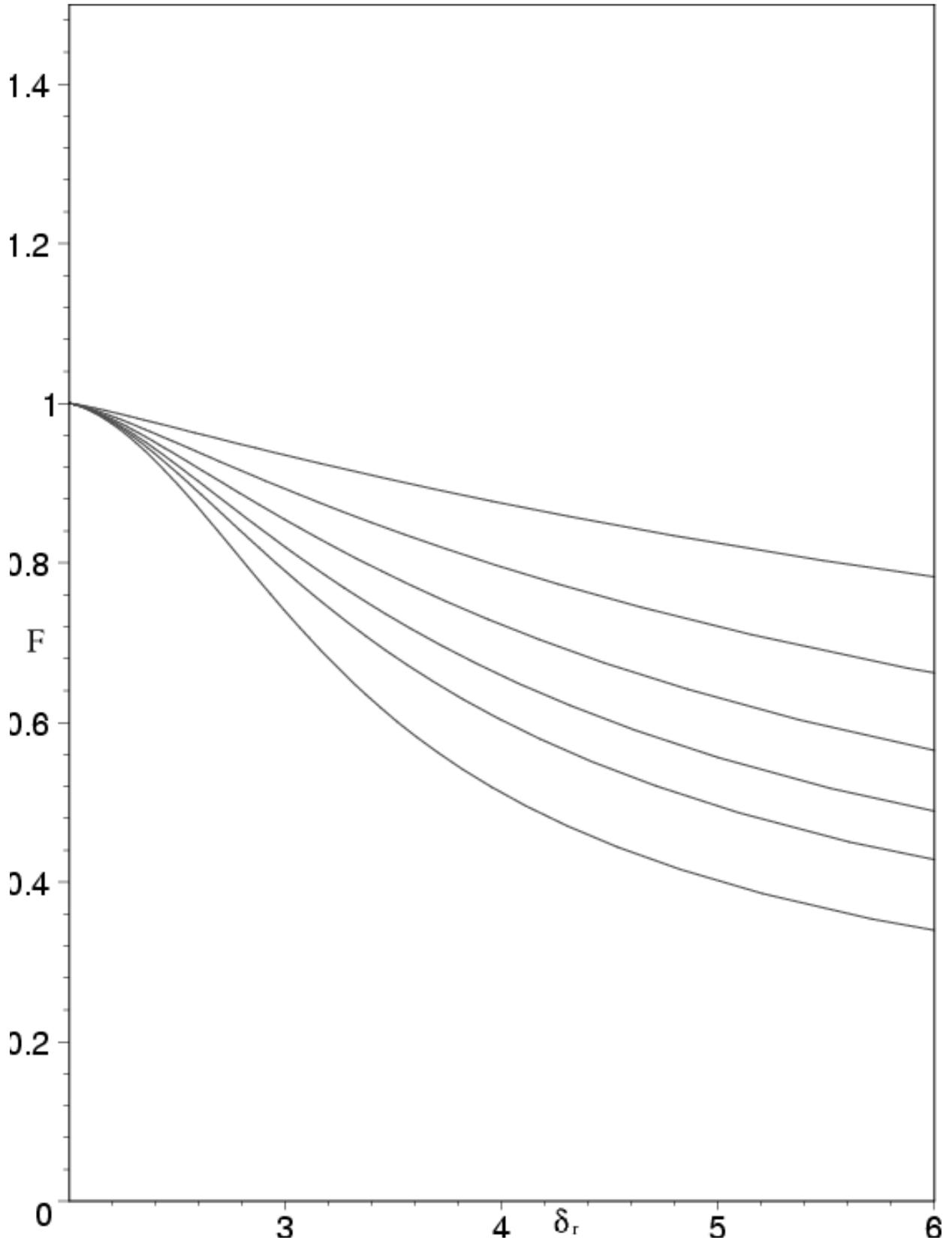}
\caption{The shock formation model in $F-\delta_r$ space for $h = 1$(top),$ 1.5, 2, 2.5, 3, 4$(bottom)}
\label{ellipsemodel}
\end{figure}

An additional constraint on the parameter space available for shock reprocessing can be obtained from studying shock formation.  A detailed study of this problem is beyond the scope of the current work, but 
a simplified treatment can provide some insight.  Fig. 
\ref{shockform} shows a schematic model of the shock forming region.  To make the problem tractable, we assume the flow forms a
shock at any point along the loop-top where the component of the flow speed normal to the loop surface is super-fast-magnetosonic  
Furthermore, we take the shock to be planar, and compute an averaged compression ratio along the shock.  
The loop geometry is taken to be elliptical, consistent with the measurements of 
\citet{A96a}.   The tangent line to the loop surface has a slope

\begin{equation}
s = \frac{x\frac{h^2}{w^2}}{\sqrt{h^2-x^2\frac{h^2}{w^2}}},
\label{slope}
\end{equation}
where $s$ is the slope of the tangent, $h$ is the loop height, $2w$ is the footpoint separation, and $x$ is the distance from the loop
center.  The component of the flow normal to the loop $v_\perp$ is given by

\begin{equation}
v_\perp = v_f \left(\frac{1}{1+s^2}\right)^{1/2},
\label{perp}
\end{equation}
where $v_f$ is the downflow speed.  Setting Eq. (\ref{perp}) equal to the fast-magnetosonic speed, $c_f$, substituting for $s$ from Eq.
(\ref{slope}), and solving for $x$ gives the critical distance, $x_{max}$, for shock formation.  The covering fraction of the shock 
is given by $F = x_{max}/w$.  The compression ratio of the shock is given by evaluating Eq. (\ref{compression}) at each point in the 
range $0 \leq x \leq x_{max}$ and averaging.
The resulting spectral index $\delta_r$ of electrons accelerated at the shock obtains from Eq. (\ref{indexshock}).  
Solutions of these equations are plotted in
Fig. \ref{ellipsemodel} for various values of $h/w$.  For all loop heights, $F$ approaches $1$ as the flow approaches the strong shock
limit of $\delta_r = 2$.  As $\delta_r$ grows large, the flow velocity decreases, and $F=0$ at the limit of $v = c_f$, or 
$\delta_r = \infty$.  Notice that as the loop gets more elongated (larger $h$), $F$ decreases.  

\begin{figure}
\epsscale{1.0}
\plottwo{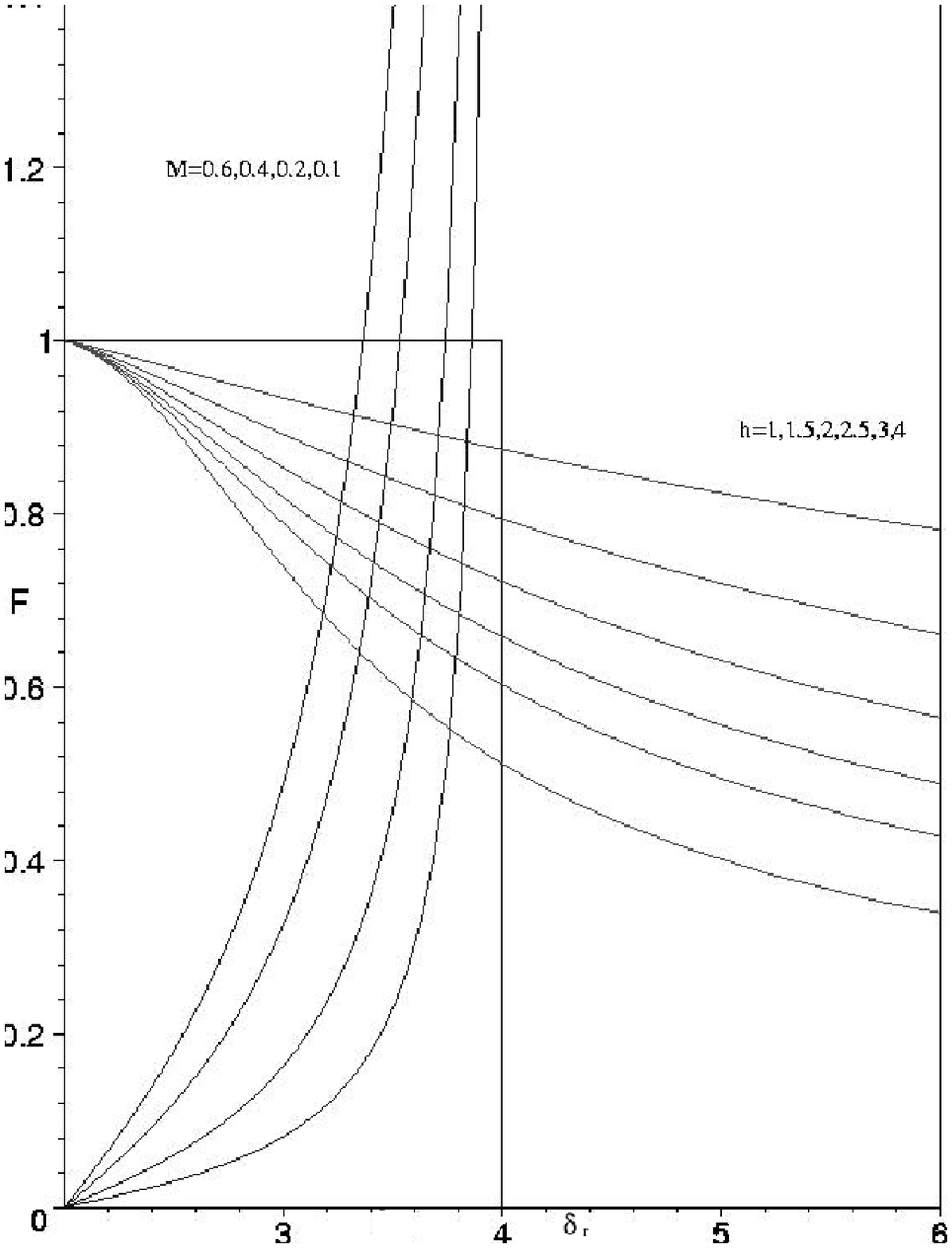}{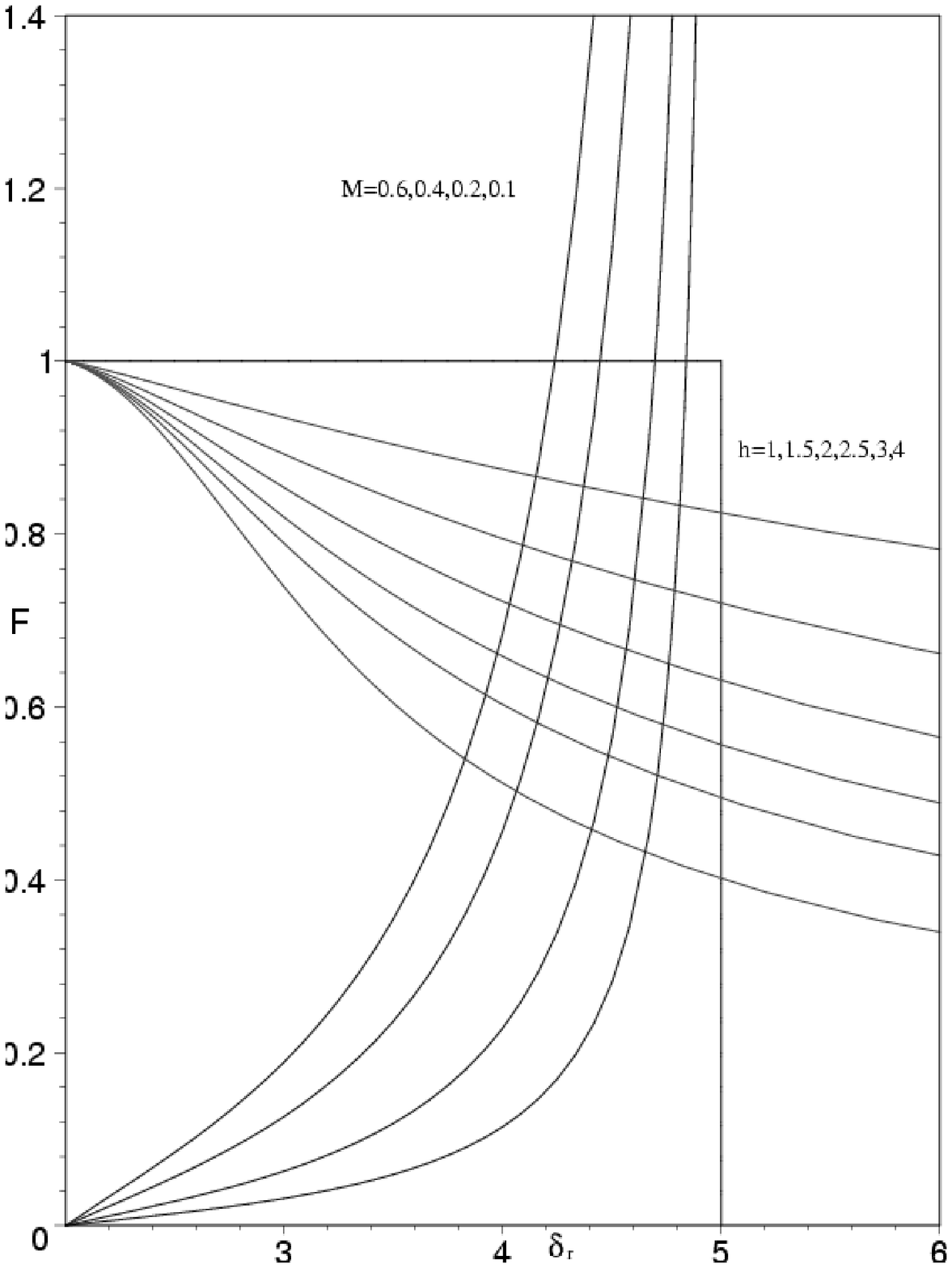}
\caption{Overlaid plots of the dual constraints of the shock reprocessing model for pulse strengths and the shock formation model.  
Left $\delta_s=4$, Right: $\delta_s=5$.  The inner box represents the physically allowed region.}
\label{crosshatchconstrain}
\end{figure}

We overlay Fig. \ref{ellipsemodel} onto panel (c) of Fig. \ref{Evariations} to produce Fig. \ref{crosshatchconstrain}, 
which incorporates the two sets
of constraints: the pulse strengths from the shock reprocessing model, and the shock formation physics.  The physically reasonable 
region, $F <1$ and $\delta_r < 4$, is delineated by the  box.  Selecting values for the two direct observables, loop height $h$ 
and pulse strength $M$ picks out a pair of curves, one from each set.  These curves have a single intersection, fixing values of 
$F$ and $\delta_r$.  Observations indicate that typically $0.1 < M <0.6$ and $h/w \backsim 1$ \citep{A96a}.
This places $3< \delta_r < 5$, consistent with the downstream fast shock spectral indexes calculated in
\citet{Blackman94} and the observed X-ray spectral indexes.

\subsection{Loop-top emission and smooth component time delays}    

An above the loop-top hard X-ray source has been observed in some impulsive phase flares, 
most notably the Masuda flare 
\citep{Masuda, Tsuneta, Sui}, but not in others.  
During the main phase of the Masuda flare, 
a kernel within the thermal source is clearly observed to be non-thermal \citep{alexander}.  
It has been argued \citep{Blackman97, Tsuneta2} that the loop-top source is associated with trapping and 
acceleration of electrons below the stationary fast shock.  In this scenario, the conditions for shock formation are met in a 
small fraction of flares, and then only marginally, resulting in weakly compressive downstream shocks in some flares, 
and no downstream shocks in others \citep{Tsuneta}.  The low shock formation rate may explain the infrequent detection of 
non-thermal loop-top emission.  It has also been argued \citep{Petrosian} that the non-thermal loop-top source
is actually prevalent in impulsive flares, but usually very dim, with typical intensity less than one tenth that of the footpoint
sources.  The contrast limit of Yohkoh is roughly $10$; observations are limited to unusually bright loop-top sources or limb 
flares with obscured footpoint emission.  We show that the shock reprocessing model is consistent with the latter scenario if STFA
takes place in the above the loop-top emission region, and is accompanied by \textit{in situ} bremsstrahlung cooling of the 
electron population.  That being said, 
there are recent RHESSI observations of two flares which are dominated by coronal emission
that is cospatial with the soft x-ray loop \citep{veronig}.  We consider this latter issue further in section 5.

Although the Masuda flare exhibits pulses, they are longer in duration
than the footpoint pulses.  
Loop-top pulses are typically of duration $\ge 10$s  \citep{tomczak}, as opposed to the sub-second
pulse structure in the footpoint emission.  
The pulsed component to which we attribute  shock reprocessing is the sub-second pulse
structure in the footpoint emission, not the much longer pulses which may come from a globally bursty
reconnection.  
When the sub-second pulses are subtracted from the emission profile 
the remaining component we refer to as the smooth profile.
In this respect, 
in addition to explaining the sporadic appearance of above the loop-top non-thermal sources, cooled STFA also produces the proper 
time delays for the smooth emission component.
\citet{A97} analyzes the arrival time delays for the smoothly varying X-ray component of 78 flares.  From these, they 
obtain plasma densities in the soft X-ray loops for 44 events, rejecting the remaining 34.  Of the rejected events, 29 were unsuitable
to the analysis technique because the fast varying component could not be sufficiently deconvolved from the smooth component.  The 
remaining 5 rejections were due to poor convergence of their fitting model.  
The remaining 44 flares have smooth component time delays consistent 
with the trap precipitation model, matching the predicted trapping time $\tau(E) \propto E^{3/2}$.  

This effect is also predicted by our shock reprocessing model.  In shock reprocessing, the pre-shock STFA 
acceleration rate is expected to determine the smooth variation time delays.  Variations in the smooth component are the result of 
changes in the STFA acceleration region, and appear at high energy on a time scale equal to the STFA acceleration time to that energy. 
The time delay curve should thus be obtained from the STFA acceleration rate:

\begin{equation}
\frac{dE}{dt} = Q(E),
\end{equation}
where Q(E) is prescribed by STFA.  We define 

\begin{equation}
\tau \equiv -\int_{E_0}^E \frac{dE}{Q(E)}.
\label{tauexpr}
\end{equation}
For non-relativistic STFA $Q(E)$ is energy independent \citep{selkblack} and  $\tau \propto E$.  For 
statistical acceleration in the quasi-linear regime, \citet{chandran} finds the standard result: $Q(E) \propto E$, and
$\tau \propto e^{E/E_0}$.  Neither of these forms is consistent with a good fit to $\tau \propto E^{3/2}$.   
Cooling in the acceleration region, which is neglected by \citet{selkblack} can be invoked to reduce the
acceleration rate.  

\begin{figure}
\epsscale{0.7}
\plotone{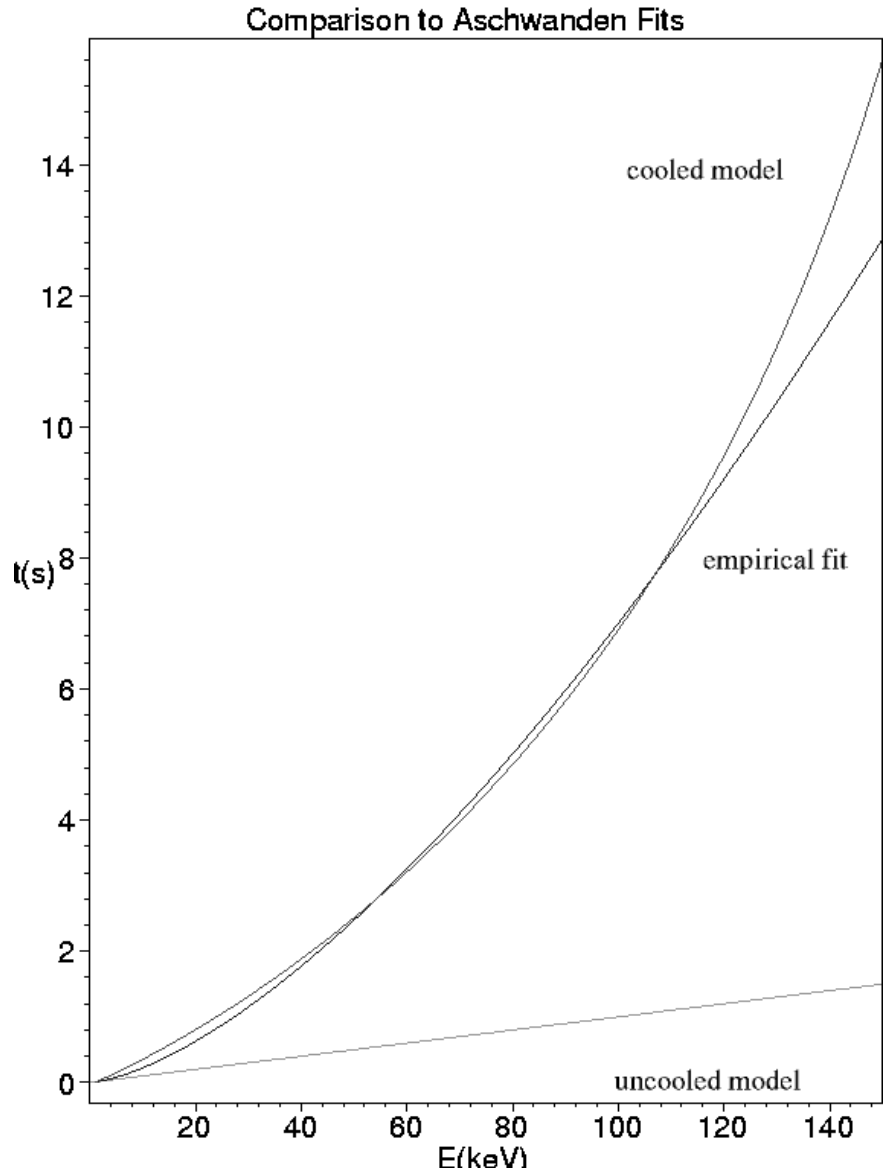}
\caption{$\tau$ vs. E for the cooled and uncooled STFA models as well as the empirical $E^{3/2}$ fit of \citet{Aschwanden}}
\label{taufig}
\end{figure}

Consider the case where STFA with \textit{in situ} cooling by bremsstrahlung occurs in a region above the loop-top shock.  
The power loss due to bremsstrahlung for a single electron is given by \citep{Rybicki}

\begin{equation}
P_B = 1.1 n_{10} E^{1/2}_1 keV s^{-1},
\end{equation}
where the dimensionless electron density $n_{10} = n/(10^{10} $cm$^{-3})$, $E_1 = E/1$keV, and we have assumed a constant Gaunt 
factor $G_{ff}= 1$.  Notice the linear density dependence for the Bremsstrahlung rate, which arises because we
consider the mean cooling rate for a single electron in a background of protons at density $n_{10}$ 
(not the emission from an ensemble
of electrons in a background of protons).  $Q(E)$ is thus given by

\begin{equation}
Q(E) = \left(\frac{dE}{dt}\right)_{STFA} - P_B = 100 n_{10}^{3/2}- 1.1 n_{10} E_{1}^{1/2},
\end{equation}
where we have adopted the STFA acceleration rate of \citet{selkblack}, which reexamined the STFA process and determined that, due to 
averaging over the limited range of pitch angles which are subject to reflection, as opposed to all pitch angles, the mean  
acceleration term dominates over the diffusive random walk component in the sub-relativistic regime. 
Taking the flare electron density to be $n_{10} = 1$ and performing the integral in equation \ref{tauexpr}  
yields the function shown in Fig. \ref{taufig} where we plot $\tau(E)$ for both cooled and uncooled STFA as well
as the empirical fit of $\tau \propto E^{3/2}$.  Cooled STFA is consistent with the time delay measurements of 
\citet{Aschwanden, A96a, A96b, A97, A98a, A98b, A99} over the observed energy range of $10-150$keV, provided we ignore collisional energy transfer to the protons.  
We follow \citet{A98b} in taking such collisional losses
to be negligible at energies above $\backsim 30$keV,
where the energy loss time is long compared to the travel time of an electron from the acceleration region to the footpoints. 
We now discuss Coulomb collisions further.

It has previously been noted \citep{Brown, schatzman} that Coulomb losses due to electron-electron scattering are typically greater
than the Bremsstrahlung cooling rate at hard x-ray energies in flare plasmas.  While this is true, within the STFA acceleration 
region, the electrons spectrum is nearly thermal.  The non-thermal post-acceleration spectrum 
results from the escape rate of 
electrons from the acceleration region \citep{selkblack} and thus the escaped electrons.  
For the distribution within the acceleration region,  Coulomb scattering among electrons does not
strongly affect a given electron's acceleration time; on average, each electron receives as much as much energy due to Coulomb scattering as it loses.  
Coulomb scattering with protons is, however, a potential cooling mechanism, since solar flare protons are 
sub-Alfv\'enic and do not participate in STFA.  The Coulomb loss rate for electron-proton scattering is given by \citep{spitzer}

\begin{equation}
\left(\frac{dE}{dt}\right)_{ep} = 2.5\times10^{-2} n_{10}E_1^{-1/2}.
\label{protoncoulomb}
\end{equation}
The electron-proton loss rate is significantly lower than the Bremmstrahlung loss rate, and electron cooling within the limited
STFA acceleration region is dominated by the latter process.  This result holds because the plasma flow participating 
in STFA contains an already nearly thermal electron population which is not interacting with cooler electrons. In the surrounding
area, which is filled with ambient plasma, STFA accelerated electrons which have escaped the acceleration region do
interact with the surrounding cooler electrons, resulting in cooling of the electrons, reshaping of the spectrum, and heating of 
the plasma.  This heated plasma may emit the thermal loop-top X-ray component observed in a number of flares.

Energy loss due to loop-top  Bremmstrahlung
cooling is significant for high energy electrons where the total energy lost to cooling is greater 
than the electron energy when it leaves the acceleration region.  The energy lost to cooling is released in x-ray luminosity.  This 
is a plausible source of the weak non-thermal loop-top emission regions observed in Yohkoh studies of flares 
\citep{Masuda, Tsuneta, Petrosian}.

\section{Discussion of observational implications and predictions}

Three major observational constraints on any model of electron acceleration in impulsive solar flares are the production of 
proper pulse strengths, the energy dependent arrival time delays, and the appearance of loop-top emission sites in only a 
fraction of observed flares.  The shock reprocessing model successfully accounts for all three of these features through a two stage 
acceleration process.  In the first phase, electrons are accelerated via STFA with appreciable thin target bremsstrahlung cooling in the
turbulent downflow region above the flare loop.  The cooling of trapped electrons produces the loop-top source, which is often
too dim relative to the footpoints to be observed by Yohkoh.  Subsequently, a portion of the electron population 
undergoes diffusive shock 
acceleration at a weakly compressive stationary fast shock, splitting the electrons into two populations: the shock reprocessed pulse
population, and the unshocked STFA population.   Sub-second pulses result from modulation of the shock strength on short time scales.  Since the 
predicted shocks are only weakly compressive, small changes in the reconnection outflow can remove or reestablish the shock.

The presence of the shock below the loop-top emission site implies that the pulse structure observed in footpoint X-rays, and generated
by variations in the shock compression ratio, are not present in loop-top sources.  This prediction of the shock reprocessing model 
needs
to be tested, and should be feasible over time as the RHESSI dataset grows to include large numbers of flares with detectable loop-top 
sources.  \citet{Sui} find evidence of a smoothly varying loop-top emission source in three RHESSI observed flares.  
There is no apparent pulse structure in these sources.  However, \citet{Sui} argue that these coronal sources are distinct from 
Masuda flare type loop-top emission sources.  The observations of \citet{tomczak} show a two component emission structure, with
both pulsed and smooth emission, superficially similar to the footpoints.  However, in these sources, the pulse time scale is typically
$10$s or larger, not sub-second as in the footpoints.  The coronal pulses are attributed to modulations in either the reconnection or 
downflow environment.  They are not inconsistent with the shock reprocessing model.    Still,
larger sampling statistics on variability in loop-top emission sources would be of great interest in determining the applicability 
of the shock reprocessing model to impulsive solar flares.

An additional concern for any model of solar flares is the electron supply problem.  For example, the hard x-rays emitted in 
the Masuda flare require a non-thermal electron throughput of $ 2 \times 10^{35}$s$^{-1}$ \citep{Nature}.  
The reconnection downflow for the same flare contained $\backsim 5 \times 10^{35}$ electrons s$^{-1}$ \citep{Tsuneta2}.  
Thus the electron acceleration mechanism must either be highly efficient, 
accelerating a fraction of the downflow electrons of order unity, or a secondary supply of electrons must be available.  STFA falls 
into the first category; the process is highly efficient, so the 
problem of supplying a sufficent flux of electrons is alleviated.  Since
the bulk plasma flows downward to the footpoints, including both the electron and proton populations, charge neutrality is maintained 
without the need to rely upon a return current.  

Shock reprocessing also makes predictions regarding the pulsed emission.  Because the pulse component of the emission is caused by
the harder shock reprocessed spectrum, the observed pulse magnitude is dependent on the energy bin, where higher energy bins show
larger pulse strengths.  From Eq. (\ref{Mpeqn}), $M \propto E_{min}^{\delta_s - \delta_r}$, where $M$ is the relative 
pulse strength, and $E_{min}$ is the minimum energy of the observations.  Notice that raising $E_{min}$ 
increases the pulse strength.  In its simplest discussed herein, 
shock reprocessing predicts that observations in a sequence of energy bins would show stronger 
pulses at higher energies.  Obtaining such data is
within the capability of current instrumentation.  The magnitude of the effect can be calculated by evaluating Eq. \ref{Mpeqn}
multiple times, with $E_{min}$ taken to be the lower energy limit of each detection band.   This provides the pulse 
strength for all x-ray emission at photon energy $E > E_{min}$.  To carry out the observational test, one canreformulate the definition of $M$ slightly by reevaluating the integrals in 
Eqs. (\ref{M_T}) and (\ref{M_p}) with the upper limit $E_{max}$ set to the high energy end of the bin.  Alternatively one may 
collectively bin all photons with energy $E > E_{min}$ for successively increasing values of $E_{min}$.  For simplicity, we perform 
the latter procedure, assuming energy bins with $E_{min}$ taken in $20$keV increments from $10$keV up to $150$keV.  For all bins, 
$E_{max} = \infty$.  The results are shown in table \ref{bintable}.  The range of pulse strengths predicted for energy bins in 
observations of a single flare is very strongly dependent on the difference $\delta_s - \delta_r$.  Even in the case of 
$\delta_r = 3.8$, a large range of pulse strengths is expected $0.07 < M < 0.54$.  
This effect can be measured, and can be used to test the shock reprocessing model and constrain $\delta_r$, 
even for moderately small values of $F$.

\begin{table}[t]
\centering
\begin{tabular}{cccc}                        
 \hline
                &                    &   $M$             &                      \\  \tableline
 $E_{min}$ (keV)&  $\delta_r=3$      & $\delta_r = 3.4$  & $\delta_r = 3.8$     \\  \tableline 
 $10$           &  $0.6$             & $0.26$            & $0.07$               \\
 $30$           &  $3.0$             & $1.1$             & $0.23$               \\
 $50$           &  $5.4$             & $1.7$             & $0.32$               \\ 
 $70$           &  $7.8$             & $2.2$             & $0.38$               \\ 
 $90$           &  $10$              & $2.6$             & $0.43$               \\ 
 $110$          &  $13$              & $3.0$             & $0.48$               \\ 
 $130$          &  $15$              & $3.4$             & $0.51$               \\ 
 $150$          &  $17$              & $3.7$             & $0.54$               \\  \hline

\end{tabular}
\caption{Pulse strengths $M$ for increasing energy bins and a range of $\delta_r$.}
\label{bintable}
\end{table}

Furthermore, the strong pulses at high energies are consistent with the soft-hard-soft (SHS) spectral pattern observed in many flare 
spectra, even down to subsecond time scales \citep{Grigis2, Grigis1}.  Typically, the SHS pattern traces total luminosity.  
As flare emission cycles through pulses, the spectrum starts out
soft at the low luminosity onset of the pulse, steadily increases up to a maximal hardness at the pulse peak, then decays back to a 
softer spectrum as the luminosity returns to the pre-pulse minimum.  The shock reprocessing model generates pulses via strengthening of 
the shock compression, which corresponds to hardening of the reprocessed spectrum, as illustrated in Table \ref{bintable}.  Small
changes in the super-magnetosonic Mach number of weakly compressive shocks can produce significant 
changes in the compression ratio.  For 
pulse strengths in excess of $1$, the footpoint X-ray spectrum will have index $\gamma_r$, the 
shock spectral index, since the emission 
will be dominated by the reprocessed population.  As the pulse progresses toward the peak luminosity, the shock compression ratio peaks
as well, and so does the spectral hardness.  During the decay phase, the shock compression decreases, and the spectrum softens, 
resulting in a SHS pattern.  Moderate pulses, especially those with large $F$ will still produce a weak SHS pattern simply due to 
the large number of reprocessed electrons.

\section{Conclusion}

Reconnection can be thought of as an environment for particle acceleration.
 Within the reconnection region and ensuing
outflows, a combination of acceleration mechanisms may operate.  
We have proposed the shock reprocessing model as one scenario of interest for solar flares
when the downflows form reconnection sites are turbulent.

Four major features of the shock reprocessing model developed in this paper are as follows:

\noindent (1) The model 
posits a two stage acceleration process: STFA in a turbulent reconnection downflow followed by first order 
acceleration at a loop-top fast shock.  Hard X-ray emission sources are predicted in the STFA region, as well as at the foot-points.  

\noindent (2) 
 Pulsed emission is produced by variations in the compression ratio of the fast shock, and thus in the post-shock electron spectrum.
The shock does not fill the entire cross section of the downflow, and is weakly compressive.  It is seen that relative pulse
strengths produced within the shock reprocessing model are reasonable for realistic values of the local plasma parameters, and a 
simple model of shock formation.  Furthermore, the generation of pulses at the shock predicts that the above the loop-top source does not
exhibit the subsecond pulse structure observed at footpoints.  

\noindent (3) 
 Both the fast (pulse) and slow (smooth) time delay measurements \citep{Aschwanden, A96a, A96b, A98a, A98b, A99} can be
explained within the shock reprocessing model.  The fast time delays in the pulsed component are time of flight dispersion
of the shock accelerated electrons.  The slow time delays in the smooth component reflect the STFA acceleration rate.

\noindent (4) 
 The shock reprocessing model predicts an increase in relative pulse strength at higher energies.  This prediction remains to be 
tested, and can distinguish between the shock reprocessing and trap precipitation models.

The shock reprocessing model is not meant to be 
an exclusive solution to electron acceleration in all solar flares.  
It seems unlikely that 
the wide variety  of flare phenomena can be explained if all reconnection regions
have the same relative contributions for the combination of acceleration processes
that can ensue.

Generic features such magnetic reconnection, soft x-ray loops, and footpoint hard x-rays, 
do appear to be common to the vast majority of flares.  
However, other phenomena are not:  Coronal x-ray emission sources appear sometimes with thermal and other times with non-thermal spectral characteristics, sometimes above the soft x-ray loop, or 
sometimes within it, and other times not at all.  
The reconnection morphology, while apparantly similar in any 2-D snapshot along the plane of the soft x-ray loops,
can have either vertical (standard) or lateral (``zipper'') dynamical evolution. 
 
One particularly difficult observation to explain within the shock reprocessing model is the appearance, in a small number of
flares, of coronal hard x-ray emission emanating from within the closed soft x-ray loop \citep{veronig}.  These sources are
non-thermal, and hard x-ray emission in the flares is dominated by the coronal, not footpoint sources.  
One might suspect such a source to be produced within the trap-precipitation model in the presence of an abnormally
dense loop; the trapped electron population within the loop would emit via thick target Bremsstrahlung.    Conversely, the shock 
reprocessing model does not load the electron beam onto the closed flare loop, and thus cannot produce these sources.  This class of
loop-top emission has only been observed in a few events.  As the total number of coronal sources within loops grows, it would be
of great interest to determine which, if any, environemntal parameters 
such as  laminar vs. turbulent downflows correlate with the location
of the coronal source.    

The phenomenological array of flares can likely be understood by  a relatively small number of acceleration
scenarios which each operate within the basic paradigm  of magnetic reconnection and outflow.  
The power source driving all flares is 
reconnection high in the corona, which launches a downflow.  Within the outflow, particle acceleration occurs, followed by
x-ray, gamma ray, and radio emission in the lower corona and chromosphere.   The particular acceleration scenario is determined by
local environmental parameters in the reconnection, downflow, and emission regions.  
Shock reprocessing and trap precipitation are two such scenarios
within this framework. A third scenario seems to 
required to explain proton dominated flares \citep{Hurford, MillerRoberts, Miller04}.

In summary: the shock reprocessing model, whereby a fraction of stochastically accelerated electrons also passes through a weakly compressive 
stationary shock as they stream toward chromospheric footpoints,
 provides a scenario for explaining a variety of features in impulsive solar flares within the more general reconnection and outflow
framework.  The model produces pulsed and smooth spectral X-ray emission components consistent with the time delay observations of 
\citet{Aschwanden, A96a, A96b, A97, A98a, A98b, A99}.  Shock reprocessing also predicts the 
appearance of above the loop-top coronal emission sites observed in a fraction of flares.  Furthermore, the model makes a sequence of 
testable predictions regarding the pulse strength as a function of energy and the SHS emission pattern.  We have discussed how it is 
observationally feasible to test the shock reprocessing model.

{\bf Acknowledgments}:
We acknowledge support from NSF grant AST-0406799, NASA
grant ATP04-0000-0016, and the KITP of UCSB, where this research was
supported in part by NSF Grant PHY-9907949.
RS acknwoledges a Horton Fellowship from the Laboratory for
Laser Energetics.

\begin{thebibliography}

 
\bibitem[Achterberg(1981)]{achterberg} 
Achterberg A.\ 1981, \aap, 98, 161 

\bibitem[Alexander \& Metcalf(1997)]{alexander} Alexander, D., \& 
Metcalf, T.~R.\ 1997, \apj, 489, 442

\bibitem[Aschwanden et al.(1999)]{A99} Aschwanden, M.~J.,
Fletcher, L., Sakao, T., Kosugi, T., \& Hudson, H.\ 1999, \apj, 517, 977

\bibitem[Aschwanden et al.(1998)]{A98a} Aschwanden, M.~J.,
Schwartz, R.~A., \& Dennis, B.~R.\ 1998, \apj, 502, 468

\bibitem[Aschwanden(1998)]{A98b} Aschwanden, M.~J.\ 1998,
\apj, 502, 455

\bibitem[Aschwanden et al.(1997)]{A97} Aschwanden, M.~J.,
Bynum, R.~M., Kosugi, T., Hudson, H.~S., \& Schwartz, R.~A.\ 1997, \apj,
487, 936

\bibitem[Aschwanden et al.(1996a)]{A96a} Aschwanden, M.~J.,
Wills, M.~J., Hudson, H.~S., Kosugi, T., \& Schwartz, R.~A.\ 1996, \apj,
468, 398

\bibitem[Aschwanden et al.(1996b)]{A96b} Aschwanden, M.~J.,
Hudson, H., Kosugi, T., \& Schwartz, R.~A.\ 1996, \apj, 464, 985

\bibitem[Aschwanden, et.al.(1995)]{Aschwanden}
  Aschwanden M.J., Schwartz R.A., Alt D.M., 1995, \apj, 447, 923 

\bibitem[Bell(1978)]{Bell}
  Bell A.R., 1978, \mnras, 182, 147

\bibitem[Bell (1978b)]{Bell2}
  Bell A.R., 1978, \mnras, 182, 443

\bibitem[Bellan(2003)]{Bellan1}
  Bellan, P.M., 2003, Advances in Space Research, 32, 1923 

\bibitem[Blackman \& Field(1994)]{Blackman94} Blackman, E.~G., \& 
Field, G.~B.\ 1994, Physical Review Letters, 73, 3097 

\bibitem[Blackman(1997)]{Blackman97}
  Blackman E.G., 1997, \apj, 484, L79


\bibitem[Blandford \& Eichler(1987)]{blandford} Blandford, R.~\& 
Eichler D.\ 1987, \physrep, 154, 1 


\bibitem[Bromund, McTiernan \& Kane(1995)]{Bromund}
 Bromund K.R., McTiernan J.M., \& Kane S.R., 1995, \apj, 455, 733

\bibitem[Brown(1971)]{Brown} Brown, J.~C.\ 1971, \solphys,
18, 489

\bibitem[Chandran(2004)]{chandran}
  Chandran B.D.G., 2004,\apj, 599, 1426


\bibitem[Chiueh \& Zweibel(1987)]{1987ApJ...317..900C} Chiueh, T., \& 
Zweibel, E.~G.\ 1987, \apj, 317, 900 
 
\bibitem[Dulk et.al. (1992)] {Dulk}
  Dulk G.A., Kiplinger A.L., Winglee R.M., 1992, , \apj, 389, 756 

\bibitem[Emslie(2003)]{emslie} Emslie, A.~G.\ 2003, \apjl, 
595, L119

\bibitem[Grigis \& Benz(2005)]{Grigis1} Grigis, P.~C., \&
Benz,
A.~O.\ 2005, \aap, 434, 1173

\bibitem[Grigis \& Benz(2005)]{NewBenz} Grigis, P.~C., \& Benz, 
A.~O.\ 2005, \apjl, 625, L143 

\bibitem[Grigis \& Benz(2004)]{Grigis2} Grigis, P.~C., \&
Benz,
A.~O.\ 2004, \aap, 426, 1093

\bibitem[Hurford et.al.(2003)]{Hurford}
Hurford G. J., Schwartz R. A., Krucker S., Lin R. P., Smith D. H., Vilmer N. 2003, \apj, 595, L77 

\bibitem[Jones \& Ellison(1991)]{jonesellison} 
  Jones F.~C.~\& Ellison D.~C.\ 1991, Space Science Reviews, 58, 259 

\bibitem[Karlick{\' y} \& Kosugi(2004)]{karlicky} 
Karlick{\' y}, M., \& Kosugi, T.\ 2004, \aap, 419, 1159 

\bibitem[LaRosa et al.(1996)]{LaRosa}
  LaRosa T.N., Moore R.J., Miller J.A., Shore S.N., 1996, \apj, 467, 454

\bibitem[Larosa \& Shore(1998)]{LarosaShore} Larosa, T.~N., \&
Shore, S.~N.\ 1998, \apj, 503, 429

\bibitem[Litvinenko(2000)]{Litvinenko}
Litvinenko, Y.~E.\ 2000, \solphys, 194, 327 

\bibitem[Masuda, et.al.(1996)]{Masuda}
  Masuda S., Kosugi T., Tsuneta S, Hara H., 1996, Adv. Space Res., 17, 63 

\bibitem[Masuda et al.(1994)]{Nature} Masuda, S., Kosugi, T.,
Hara, H., Tsuneta, S., \& Ogawara, Y.\ 1994, \nat, 371, 495

\bibitem[Melrose(1974)]{melrose}
  Melrose D.B., 1974, \solphys, 37, 353

\bibitem[Melrose \& Brown(1976)]{BrownMelrose} Melrose, D.~B., \&
Brown, J.~C.\ 1976, \mnras, 176, 15


\bibitem[Miller, Emslie, and Brown(2004)]{Miller04}
  Miller J.A., Emslie A.G., Brown J.C., 2004, \apj, 602, L69

\bibitem[Miller, Larosa, \& Moore(1996)]{Miller} 
Miller J.~A., Larosa T.~N., Moore R.~L.\ 1996, \apj, 461, 445

\bibitem[Miller and Roberts(1995)]{MillerRoberts}
Miller J.A., Roberts D.A., 1995, \apj, 452, 912

\bibitem[Mrozek \& Tomczak(2004)]{tomczak} Mrozek, T., \& 
Tomczak, M.\ 2004, \aap, 415, 377 

\bibitem[Petrosian et al.(2002)]{Petrosian} Petrosian, V.,
Donaghy, T.~Q., \& McTiernan, J.~M.\ 2002, \apj, 569, 459

\bibitem[Priest \& Forbes(2002)]{Priest} 
Priest, E.~R., \& Forbes, T.~G.\ 2002, \aapr, 10, 313 

\bibitem[Rybicki \& Lightman(1979)]{Rybicki} Rybicki, G.~B., \&
Lightman, A.~P.\ 1979, New York, Wiley-Interscience, 1979.~393 p.,

\bibitem[Schatzman(1965)]{schatzman} Schatzman, E.\ 1965, ASSL 
Vol.~1: The Solar Spectrum, 313 

\bibitem[Selkowitz \& Blackman(2004)]{selkblack} Selkowitz, R., 
\& Blackman, E.~G.\ 2004, \mnras, 354, 870 

\bibitem[Somov \& Kosugi(1997)]{Kosugi} 
Somov, B.~V., \& Kosugi, T.\ 1997, \apj, 485, 859 

\bibitem[Spitzer(1956)]{spitzer}
  Spitzer L., 1956, Physics of Fully Ionized Gases. Interscience, New York

\bibitem[Sui et al.(2004)]{Sui} Sui, L., Holman, G.~D., \&
Dennis, B.~R.\ 2004, \apj, 612, 546

\bibitem[Tsuneta(1996)]{Tsuneta}
  Tsuneta S., 1996, \apj, 456, 840

\bibitem[Tsuneta \& Naito(1998)]{Tsuneta2} Tsuneta, S., \& 
Naito, T.\ 1998, \apjl, 495, L67 

\bibitem[Veronig \& Brown(2004)]{veronig} Veronig, A.~M., \& 
Brown, J.~C.\ 2004, \apjl, 603, L117

\end{thebibliography}
\end{document}